\documentclass[%
 aip,amsmath,amssymb,reprint,superscriptaddress
]{revtex4-1}

\usepackage{graphicx}
\usepackage{dcolumn}
\usepackage{bm}
\usepackage{enumerate}
\usepackage{eurosym}
\usepackage{amsfonts}
\usepackage{amsmath}
\usepackage{amsopn}
\usepackage{amssymb}
\usepackage{graphicx}
\usepackage{graphics}
\usepackage[colorlinks=true, allcolors=blue]{hyperref}
\usepackage{color}
\usepackage{verbatim}
\usepackage{mathrsfs}
\usepackage{hyperref}
\usepackage{braket}
\usepackage{graphicx}% Include figure files
\usepackage{dcolumn}% Align table columns on decimal 

\makeatletter
\def\@email#1#2{%
 \endgroup
 \patchcmd{\titleblock@produce}
  {\frontmatter@RRAPformat}
  {\frontmatter@RRAPformat{\produce@RRAP{*#1\href{mailto:#2}{#2}}}\frontmatter@RRAPformat}
  {}{}
}%
\makeatother
\begin{document}

\preprint{AIP/123-QED}

\title[Quantum Quasinormal Mode Theory for Dissipative Nano-Optics and Magnetodielectric Cavity Quantum Electrodynamics]{Quantum Quasinormal Mode Theory for Dissipative Nano-Optics and Magnetodielectric Cavity Quantum Electrodynamics}
%%%%%%%%%%%%%%%%%%%%%%%%%%%
\author{Lars Meschede}
\affiliation{School of Physics and CRANN Institute, Trinity College Dublin, The University of Dublin, College Green, Dublin 2, Ireland}
%%%%%%%%%%%%%%%%%%%%%%%%%%%
\author{Daniel D. A. Clarke}
\affiliation{School of Physics and CRANN Institute, Trinity College Dublin, The University of Dublin, College Green, Dublin 2, Ireland}
\email{clarked5@tcd.ie}
%%%%%%%%%%%%%%%%%%%%%%%%%%%
\author{Ortwin Hess}
\affiliation{School of Physics and CRANN Institute, Trinity College Dublin, The University of Dublin, College Green, Dublin 2, Ireland}
%\affiliation{AMBER, SFI Research Centre for Advanced Materials and BioEngineering Research, Trinity College Dublin, the University of Dublin, College Green, Dublin 2, Ireland}
\email{ortwin.hess@tcd.ie}

\date{\today}% It is always \today, today,

\begin{abstract}
    The unprecedented pace of evolution in nanoscale architectures for cavity quantum electrodynamics (cQED) has posed crucial challenges for theory, where the quantum dynamics arising from the non-perturbative dressing of matter by cavity electric and magnetic fields, as well as the fundamentally non-hermitian character of the system are to be treated without significant approximation. The lossy electromagnetic resonances of photonic, plasmonic or magnonic nanostructures are described as quasinormal modes (QNMs), whose properties and interactions with quantum emitters and spin qubits are central to the understanding of dissipative nano-optics and magnetodielectric cQED. Despite recent advancements toward a fully quantum framework for QNMs, a general and universally accepted approach to QNM quantization for arbitrary linear media remains elusive. In this work, we introduce a unified theoretical framework, based on macroscopic QED and complex coordinate transformations, that achieves QNM quantization for a wide class of spatially inhomogeneous, dissipative and dispersive, linear, magnetodielectric resonators. The complex coordinate transformations equivalently convert the radiative losses into non-radiative material dissipation, and via a suitable transformation that reflects all the losses of the resonator, we define creation and annihilation operators that allow the construction of modal Fock states for the joint excitations of field-dressed matter. By directly addressing the intricacies of modal loss in a fully quantum theory of magnetodielectric cQED, our approach enables the exploration of modern, quantum nano-optical experiments utilizing dielectric, plasmonic, magnetic or hybrid cQED architectures, and paves the way towards a rigorous assessment of room-temperature, quantum nanophotonic technologies without recourse to {\it ad hoc} quantization schemes.
\end{abstract}

\maketitle

\section{Introduction}

Cavity-mediated control and enhancement of light-matter interaction has become a pervasive paradigm in modern quantum science and technology. Already decades in maturity, the use of micro- and nanoscale, optical, dielectric resonators to both harness and manipulate the photonic properties of semiconductor quantum dots has enabled the exploration of fundamental cavity quantum electrodynamics (cQED) in the solid state~\cite{vahala2003optical,khitrova2006vacuum,kirakochbook}, as well as a host of device applications ranging from high-performance, non-classical light sources and spin-photon interfaces to ultrafast, all-optical switching for integrated photonic quantum computation and networks~\cite{hepp2019qd,arakawaholmes2020progress,heindel2023qd}.
%photon-photon gates to spin-based quantum computation~\cite{michlerbook}.
In stark contrast to their dielectric counterparts, plasmonic nanocavities offer the unique ability to confine light to extremely sub-wavelength scales and massively enhance local electromagnetic fields via their resonant surface plasmon modes, thereby constituting unique architectures for the exploration of light-matter interaction free from diffraction bounds~\cite{baumberg2019extreme}. Crucially, plasmonic resonators have ushered in an unprecedented era of room-temperature cQED at the nanoscale, facilitating access to the regime of strong coupling even at the profound limit of single molecular and quantum dot emitters under ambient conditions~\cite{chikkaraddy2016single,santhosh2016vacuum,gross2018near,park2019tip}, and opening up a compelling vista of potential applications in quantum nanotechnologies~\cite{xiong2021room,clarke2024near}, extending from single-qubit coherent control~\cite{zhang2017sub} and ultrafast single-photon emission~\cite{hoang2016ultrafast,bogdanov2020ultrafast,you2020reconfigurable} to near-field multipartite entanglement generation~\cite{bello2020controlled,bello2022near} and quantum sensing~\cite{kongsuwan2019sensing}.
%Complementary to dielectric and plasmonic architectures, where cQED effects in dielectric and plasmonic are mediated by the electric dipole coupling between light and matter, magnonic cQED designs that leverage ferro- or ferrimagnetic nanostructures
%resonance in magnetic nanostructures has a recent surge of interest in the magnetic dipole coupling between
In contrast to the electric dipole coupling between light and matter, cQED phenomena emerging from magnetic interactions have also been the subject of increasing attention. Leveraging collective microwave excitations of the spins in ferromagnetic and ferrimagnetic structures, the demonstration of such effects as photon-magnon and cavity-mediated magnon-qubit strong coupling~\cite{zhang2014strongly,hou2019strongmagnon,li2019magnon,neuman2020nanomagnonic},
%as well as magnon-mediated spin-spin interactions~\cite{neuman2020nanomagnonic}
have established cavity (nano-)magnonics as an attractive paradigm~\cite{rameshti2022cavity}, with interesting prospective applications in the engineering of synthetic gauge fields~\cite{gardin2024engineering}, cavity-mediated spin-photon interfaces at the nanoscale~\cite{chen2024scalable}, and the realization of quantum networks of entangled spins.
%designing quantum state manipulation, transduction and non-reciprocal networking protocols.
%On the other hand, also phenomena originating from the magnetic component of the electromagnetic interaction have received increased attention, such as photon-magnon and cavity-mediated magnon-qubit strong coupling in cavity magnonics~\cite{rameshti2022cavity}, with interesting perspective applications in quantum transduction and engineering of synthetic gauge fields~\cite{gardin2024engineering}, or cavity-mediated spin-photon interfaces at the nanoscale~\cite{chen2024scalable}, offering a possibility for the realization of quantum networks of entangled spins.

As duly highlighted in the literature, the ability to describe the optical response of photonic, plasmonic or magnonic resonators and their interaction with quantum emitters (QEs) and spin qubits in terms of a small number of modes is of significant practical utility. A cavity mode formalism renders more transparent the physical attributes of light-matter interaction, allows one to isolate and quantify both radiative and non-radiative contributions to the observed dynamics, and aids device optimization towards particular applications. For ideal, closed cavities in the absence of material absorption (\textit{i.e.}, hermitian systems), normal modes emerge with real eigenfrequencies (resonances), infinite lifetimes and which are orthonormal with respect to some scalar product. In practice, the dielectric microcavities of traditional semiconductor quantum optics (which feature very high quality factors) most closely approximate this scenario~\cite{vahala2003optical,khitrova2006vacuum}, but nevertheless, the notion remains ambiguous in general and cannot be applied without further considerations~\cite{kristensen2012mode,kristensen2014modes}. The issue of decay is especially important for nanoplasmonic resonators, where Ohmic dissipation inherent to their metallic constituents and open-cavity radiation losses are essentially universal, but has also garnered interest in the context of magnonic cavity systems where dissipative coupling effects have been observed~\cite{wang2020dissipativemag,yu2024mag} Of course, these losses typically have a deleterious impact on the performance of schemes for quantum information processing~\cite{xiong2021room}, but could also be harnessed in order to engineer the functionality of quantum metamaterials and to promote novel veins of research in non-hermitian topological photonics or in the quantum thermodynamic facets of nanophotonic technologies. In all such cases, it is evidently desirable to capture in full the inherently dissipative character of the electromagnetic resonances.

To date, one of the most powerful and increasingly widespread approaches to this end has been quasinormal mode (QNM) theory~\cite{lai1990time,leung1994completeness,leung1994perturbation,leung1996dielectric,lee1999dyadic,kristensen2014modes,yan2018rigorous,lalanne2018light,sauvan2022norm}. The QNMs are time-harmonic solutions of Maxwell's equations, satisfying suitable outgoing-wave boundary conditions (\textit{e.g.}, the Silver-M{\"u}ller radiation condition) and bearing complex eigenfrequencies, loss thus being accounted for as an intrinsic modal property. The imposition of a radiation condition ensures that light propagates away from the cavity (as expected for a leaky resonator), but in conjunction with the complex nature of their eigenfrequencies, gives rise to a mathematically and conceptually difficult property of QNMs, namely their exponential divergence at large spatial distances from the resonator. More generally, the non-conservative nature of lossy cavity systems has raised a variety of non-trivial questions regarding a proper, mathematically rigorous formulation of normalization and orthogonality, alongside the matter of whether or not the QNMs truly form a complete set, and thus allow mode expansions for arbitrary electromagnetic field distributions.
%Such issues also pose a challenging obstacle for the development of a quantum-optical theory as well,
In spite of the challenges and ambiguities, classical QNM methodology has undergone significant advances in the last decade, including improved understanding of the formal aspects of the theory~\cite{sauvan2022norm},
%(mode normalization, orthogonality and completeness of the QNMs)
techniques for computing
%regularized (divergence-free)
the QNMs of arbitrary resonators~\cite{bai2013efficient,sauvan2013theory,ge2014time,ge2014quasinormal,yan2018rigorous,lalanne2018light} and the release of general-purpose software~\cite{wu2023man}, as well as successful applications to the calculation of generalized effective mode volumes~\cite{kristensen2012mode,sauvan2013theory,kristensen2014modes}, electromagnetic near- and far-field characteristics~\cite{kongsuwan2020qnm,ren2020nfft,bedingfield2023multi}, Green's functions~\cite{lee1999dyadic,ge2014quasinormal}, enhanced spontaneous emission rates~\cite{sauvan2013theory,ge2014quasinormal,dezfouli2017modal,ren2024classical}, the semi-classical treatment of resonator-QE systems in the weak coupling regime~\cite{yang2015analytical}, and even elastic Purcell factors in optomechanical systems~\cite{elsayed2020elastic}.

Inevitably, a quantum description of lossy electromagnetic modes is essential towards the fundamental study and understanding of quantum nanophotonic systems, including the statistics of photon, plasmon or magnon emission, polaritonic effects and the role of quantum fluctuations, as well as to facilitate rigorous quantification of key figures of merit for emerging quantum nano-optical technologies.
%Notwithstanding the marked progress in fundamental QNM research over the years, many of the aforementioned works have not been conducive to a rigorous quantization protocol for QNMs and their interaction with quantum matter.
To date, most theoretical investigations of nanoscale cQED and spectroscopy (\textit{e.g.}, surface-enhanced and electron energy loss spectroscopies) have relied on an {\it ad hoc} quantization protocol, in which the bosonic cavity mode is directly modeled as a quantum harmonic oscillator without formal, field-theoretic justification, while decay effects are incorporated via the relevant dissipators in a quantum master equation, themselves derived from a phenomenological coupling to a bath of continuum modes~\cite{neuman2020sers,digiulio2019probing,zhou2022shb,hummer2013weak}. Such a treatment is expected to work well if the resonator response is dominated by a single mode or if the spectral density has a simple, multi-Lorentzian profile in the multimode case, as in Ref.~\cite{crookes2025collective} for example, but breaks down if the spectral density exhibits a more complex structure.
%elementary photon/plasmon absorption and emission processes, hybridization,......photon/plasmon statistics, rigorous quantification of performance metrics for emerging quantum nanophotonic devices demands a quantum description of their lossy electromganetic resonances,
Recently however, significant progress was made with the introduction of a second quantization scheme for QNMs, enabling the construction of multiplasmon or multiphoton Fock states for arbitrary, three-dimensional (3D), dissipative and non-magnetic resonators~\cite{franke2019quantization}. This approach has been elaborated and applied to a range of problems, such as concerning the limits of (phenomenological) dissipative Jaynes-Cummings models in plasmonic cQED, the performance of near-field-driven single-photon sources~\cite{hughes2019spe}, the description of nonlinear cQED effects in hybrid metal-dielectric cavity systems~\cite{franke2020nonlinear} and coupled lossy and amplifying resonators~\cite{franke2022coupled}. An important feature of these works is the regularization of the QNMs based on a Dyson equation formalism
%; the regularized modes obtained in this way agree with the QNMs inside the resonator, but behave as functions of real frequency for positions outside the resonator.
and a dual expansion scheme for the dyadic Green's function~\cite{ge2014quasinormal}, which utilizes the unregularized modes for positions inside the resonator and regularized modes that are functions of real frequency for positions outside.
%in which the QNMs are replaced by regularized modes that agree with the QNMs inside the resonator and are functions of real frequency for positions outside the resonator.
Whilst the Dyson equation approach appears to yield an accurate representation for the Green's function outside the resonator volume and allows accurate calculation of the Purcell factor in this region for the studied systems~\cite{ge2014quasinormal}, numerical computation of the regularized fields appears computationally intensive, and may even become intractable for more complex photonic cavity systems. Moreover, the formal justification for such a mode regularization technique remains elusive~\cite{sauvan2022norm}, and in particular, is based on an assumed completeness of the QNMs inside the resonator. Whilst this may apply to simple geometries (\textit{i.e.}, compact resonators in a homogeneous background medium), it is not guaranteed for more complex geometries, such as resonators on a substrate or waveguide-coupled resonator systems~\cite{sauvan2022norm}. Furthermore, for non-trivial backgrounds, knowledge of the corresponding background Green's function would be required for regularization, the determination of which could itself be non-trivial.
%\textcolor{red}{E-field operator is expanded in unregularized QNMs, not valid for positions farer away!}
%Moreover, the formal justification for such a mode regularization technique remains elusive~\cite{sauvan2022norm}.

Alternative quantization schemes for lossy and non-magnetic resonators have also emerged in the last few years, relying on a fitting of the spectral density of the electromagnetic environment~\cite{medina2021few, sanchez2022few} based on emitter-centered modes~\cite{buhmann2008casimir,feist2020macroscopic,hummer2013weak,rousseaux2016adiabatic}, or a transformation of the continuum of photonic eigenmodes into a discrete set of pseudomodes~\cite{yuen2024exact}.
The former offers a practical method to describe the interaction of QEs with the electromagnetic environment in terms of a few interacting lossy modes, and also requires knowledge of the photonic Green's function. These modes, however, do not have a direct physical interpretation and their number is not fixed (scaling with the number of QEs featuring in the problem), while it remains unclear if the electric field can be
%The authors attributed the discrete modes a field profile based on the emitter-centered modes; however, to the best of our knowledge, they did not check if the E-field could be
correctly recovered from these discrete modes alone, as would be desirable for a modal theory. Nevertheless, fewer of these modes may be needed relative to the QNM case, which may become especially relevant in the vicinity of accumulation points of QNMs (\textit{e.g.}, due to high-wavevector plasmonic resonances). In contrast, the pseudomode theory offers the notable advantage of circumventing traditional reservoir approximations and can thereby capture non-Markovian effects in the dynamics of nanocavity-QE systems. Notably though, the theory is formulated on the basis of conventional QED, by assuming that the electric field can be expanded in terms of a continuum of normal modes, and consequently cannot be used to describe lossy or dispersive systems. Furthermore, it remains unclear how efficiently the method can be applied to systems that do not admit analytical tractability (in particular, for morphologically complex resonators), and if it can be used for arbitrary initial conditions beyond those pertinent to the single-excitation regime.

Notwithstanding the aforementioned efforts, it remains desirable to formulate a unified and numerically feasible methodology for describing the lossy resonances of arbitrary, linear, magnetodielectric systems, directly addressing such crucial issues as mode divergence, the identification of canonical field variables in the presence of both dielectric and magnetic media, and the calculation of key figures of merit for quantum nano-optical device applications on rigorous foundations. It is worth mentioning here that the quantization of QNMs and their coupling to QEs is similar to the reaction coordinate mapping of a (strongly-coupled) spin-boson model with a structured bath, described by a spectral density peaked around a certain frequency~\cite{binder2018thermodynamics,anto2021capturing}. In this method, the bosonic continuum is decomposed into a collective bosonic degree of freedom and a residual continuum, for which the interaction can be treated within the Markov approximation. 
%quantum dynamics arising from the non-perturbative dressing of matter by cavity electric and magnetic fields as well as the fundamentally non-hermitian character of the system are to be treated without significant approximation.
%based on macroscopic QED remains desirable.
%\textcolor{red}{Pseudomode quantization does not start from Green's function approach (no noise current added)?}
%the development of efficient near-field-to-far-field transformations

In this work, we present a unified theoretical framework that achieves QNM quantization for a wide class of 3D, spatially inhomogeneous, dissipative (with possible gain components) and dispersive, linear, magnetodielectric resonators. Our approach is based on a rigorously defined mode regularization and truncation of the spatial domain
via exterior complex coordinate transformations, tantamount to perfectly matched layers (PMLs)~\cite{hugonin2005pml,sauvan2022norm,demesy2023dispersive}.
%For numerical computations,
Harnessing established auxiliary-field eigenvalue techniques (or suitable generalizations thereof) with PMLs,
%to obtain
a complete set of regularized modes can be ascertained, enabling the deployment of modal expansions for the dyadic Green's function and electromagnetic fields both inside and outside the resonator volume.
The quantization of these modes is then treated in the formalism of macroscopic QED (mQED), facilitating the construction of discrete Fock states for the joint excitations of field-dressed matter
%dressed by cavity electric and magnetic fields,
in a manner that fully accounts for
%in the presence of
both radiative and non-radiative modal losses, and which provides the foundations for a quantum dynamic theory via quantum Langevin and master equations.
%By basing our quantum description of QNMs on a rigorously defined regularization procedure,
Our present methodology attains unprecedented scope among quantum QNM techniques, offering a consolidated approach to the analysis of contemporary quantum nano-optical experiments utilizing dielectric, plasmonic, magnonic or hybrid cQED architectures, and paves the way towards a
%Moreover, our fully quantized treatment enables a
rigorous and numerically feasible assessment of the performance metrics for emerging, room-temperature-viable, quantum nanophotonic technologies without recourse to
%ranging from near-field-driven strong coupling and multipartite entanglement generation and non-linear cQED systems.
%avoiding reliance on {\it ad hoc} quantization schemes.
%Crucially, our fully quantized framework enables a rigorous assessment of performance metrics for emerging, room-temperature-viable quantum nanophotonic technologies, avoiding reliance on
{\it ad hoc} quantization schemes.
%Leveraging efficient QNM eigensolvers and the well-established PML methodology, our approach provides a computationally efficient and broadly applicable  dielectric, plasmonic, magnonic and hybrid cQED architectures.    

The paper is organized as follows.
\S~\ref{Sec:Classic_QNMs}
%presents the general theory of PML-based QNM quantization. In particular, \S~\ref{Sec:Classic_QNMs}
briefly reviews the classical electromagnetic QNM theory and PML regularization, as well as the auxiliary-field eigenvalue technique that yields a complete set of orthonormalized modes, alongside useful formulae for time-domain analysis, in the commonly treated case of non-dispersive magnetic permeabilities. In \S~\ref{Sec:Quantum_QNMs}, we discuss the theory of mQED in arbitrary, linear, magnetodielectric media; here, we quantize the modes by incorporating PML regularization in the formalism and defining bosonic modal operators in order to construct the corresponding Fock space. Using these operators, we derive an effective Lindblad master equation in \S~\ref{Sec:Master_Eq} to describe the dynamics of the quantized modes and their coupling to QEs. In \S~\ref{Sec:Numerical_Examples}, we demonstrate the theory by investigating the spontaneous decay of a QE in a simple, 1D, half-open cavity (\S~\ref{Sec:1D_example}) and in a 3D, spherical cavity (\S~\ref{Sec:3D_example}), comparing our predictions with the semi-analytical results of an exact mQED treatment. Finally, we close the paper in \S~\ref{Sec:Conclusion} with some concluding remarks and perspectives.

%This section presents a general mode theory for QNMs using complex coordinate transformations. In \S~\ref{Sec:Classic_QNMs} we present the general ideas behind the mode regularization, normalization, and completeness, and in \S~\ref{Sec:Quantum_QNMs} we present the quantization of the QNMs. 

%\textcolor{red}{I would delete this section and make the subsections below all their own sections and summarize their content in the introduction.}

\section{\label{Sec:Classic_QNMs} Classical QNMs: Regularization, Normalization and Completeness Based on Stretched Coordinates}

%\subsection{\label{subsec:Classical_QNMs_general} Modal formulation of the scattering problem}

In this section, we review the classical electromagnetic QNM theory based on PML regularization, as introduced in Refs.~\cite{sauvan2013theory, vial2014quasimodal,yan2018rigorous} and reviewed in Ref.~\cite{lalanne2018light}. Our aim is to provide a reasonably self-contained presentation that captures the key concepts and important formulae that are also required in the quantized treatment, whilst additionally establishing the relevant notation.

\begin{figure}
    \centering
    \includegraphics[width=1\linewidth]{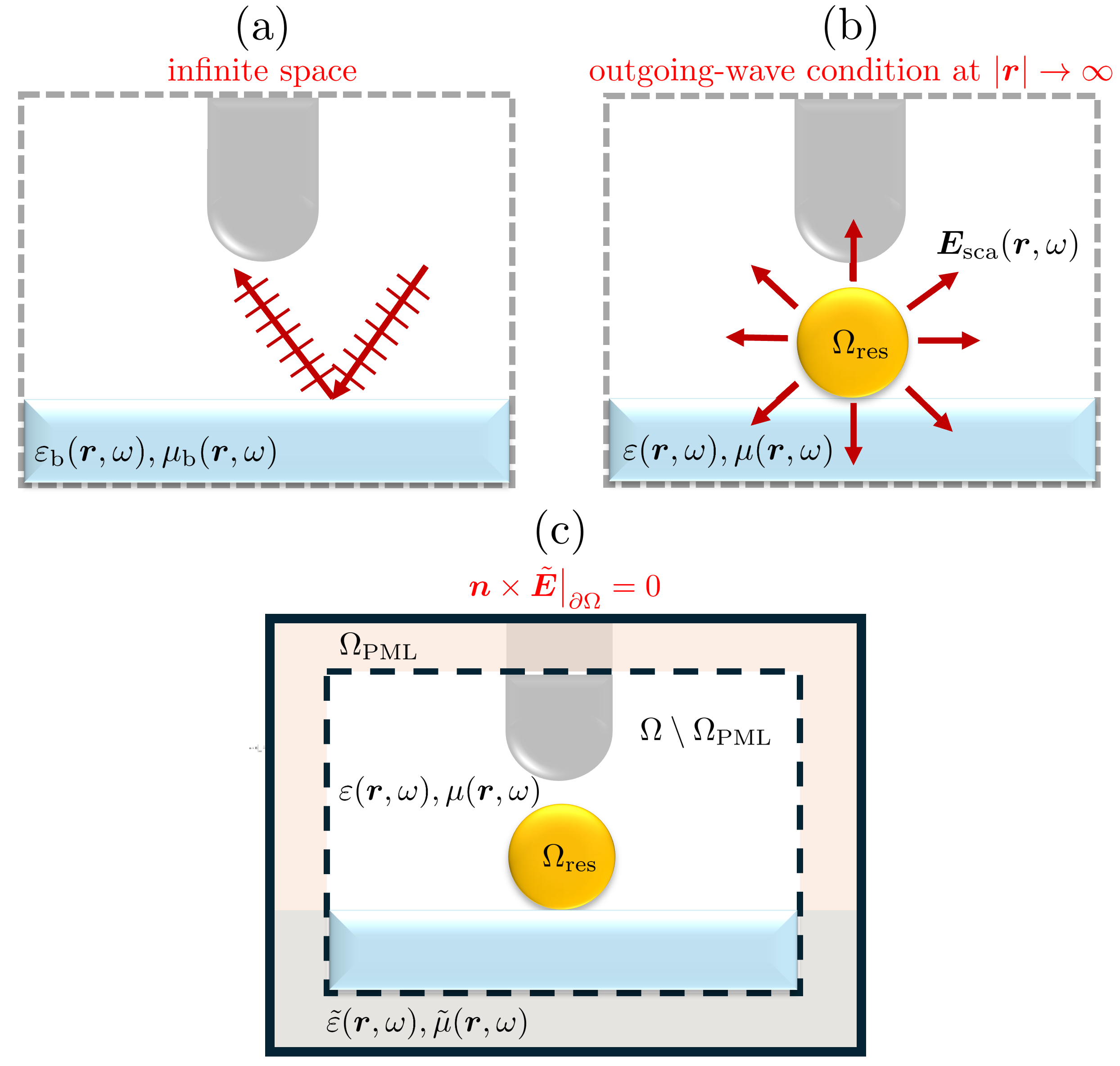}
    \caption{Concept of the background-field formulation and the PML-based truncation of the domain used for QNM calculation, regularization and quantization. (a) Schematic of an example background, described by material parameters $(\varepsilon_\mathrm{b}, \mu_\mathrm{b})$, in which Maxwell's equations have to be satisfied by the background/incident field $\bm{E}_\mathrm{b}$ in the infinite space. The background can be non-trivial and contain substrates or waveguides. (b) The scattered-field problem in which Maxwell's equations have to be satisfied by $\bm{E}_\mathrm{sca}$ in the full geometry, including the resonator with compact domain $\Omega_\mathrm{res}$ and containing the effective current density arising from $\bm{E}_\mathrm{b}$. Suitable outgoing-wave conditions in the infinite space are applied. (c) The corresponding problem for $\bm{E}_\mathrm{sca}$ with PMLs imposed, in which the infinite space is truncated to a compact domain $\Omega$. Using a complex coordinate transformation, which leaves the inner domain $\Omega \setminus \Omega_\mathrm{PML}$ invariant and causes outgoing waves to be damped in $\Omega_\mathrm{PML}$, is equivalent to the introduction of an effective material $(\tilde{\varepsilon}, \tilde{\mu})$. The damping of the outgoing waves effectively converts the outgoing-wave condition to a Dirichlet condition (perfect electrical conductor) at infinity, and due to the exponential nature of the damping, this boundary condition can very well be approximated using a finite truncated domain $\Omega$ and imposing the condition at $\partial\Omega$.}
    \label{Fig_PML}
\end{figure}

% In general, the electromagnetic resonances of a 3D, lossy and dispersive resonator are most rigorously treated as QNMs~\cite{sauvan2022norm}.
We consider a system comprising a lossy resonator, contained in some compact region $\Omega_\mathrm{res} \subset \mathbb{R}^3$ and embedded in some arbitrary linear material background described by a relative permittivity $\varepsilon_\mathrm{b}(\bm{r},\omega)$ and relative permeability $\mu_\mathrm{b}(\bm{r},\omega)$, while we denote the relative material parameters of the whole system by $\varepsilon(\bm{r},\omega)$ and $\mu(\bm{r},\omega)$. In general, the background does not need to be homogeneous and could, for example, contain a substrate or waveguides, as schematized in Fig.~\ref{Fig_PML}(a). The QNM electric and magnetic fields, ${\bm E}_n$ and ${\bm H}_n$ respectively, are defined as the time-harmonic solutions of the source-free Maxwell's equations,
%at position $\bm{r}$ and perturbation frequency $\omega$ are described by electric permittivity $\varepsilon(\bm{r},\omega)$ and magnetic permeability $\kappa=\mu(\bm{r},\omega)^{-1}$,
\begin{align}\label{classicalQNM}
    \begin{bmatrix}
    0 & -\mathrm{i}\mu_0^{-1}{\mu}^{-1}(\bm{r},\tilde{\omega}_n)\nabla\times \\
    \mathrm{i}\varepsilon_0^{-1}{\varepsilon}^{-1}(\bm{r},{\tilde \omega}_n)\nabla\times & 0 \\
    \end{bmatrix}
    \begin{bmatrix}
    \bm{H}_n \\
    \bm{E}_n
    \end{bmatrix}\notag \\
    = {\tilde \omega}_n
    \begin{bmatrix}
    \bm{H}_n \\
    \bm{E}_n
    \end{bmatrix},
\end{align}
subject to an appropriate outgoing-wave boundary condition (note that we have assumed a time dependence of the form $\mathrm{e}^{-\mathrm{i}\omega t}$). Alternatively, the QNM problem in Eq.~\eqref{classicalQNM} can be formulated as an equation for the electric field only:
\begin{align}
    \nabla \times \left( \mu^{-1}(\bm{r},\tilde{\omega}_n) \nabla \times \bm{E}_n\right) - \frac{\tilde{\omega}_n^2}{c^2} \varepsilon(\bm{r},\tilde{\omega}_n)\bm{E}_n = 0 \ . 
\end{align}
In the case of a homogeneous background, one can use the Silver-M{\"u}ller radiation condition, but in a more complex background the condition needs to be adjusted~(see, for example, Refs. \cite{kristensen2014calculation,kristensen2017theory}).
%Throughout, we assume an $e^{-i\omega t}$ time-dependence for the electric and magnetic fields.
%In Eq. (\ref{classicalQNM}), $\varepsilon(\bm{r},\omega)$ and $\mu(\bm{r},\omega)$ denote the electric permittivity and magnetic permeability distributions of the system, which may depend on both position ${\bm r}$ and frequency $\omega$
%The QNM eigenfrequencies are necessarily complex in character, ${\tilde{\omega}}_{n} = \omega_{n} - i\Gamma_{n}/2$, with real and imaginary parts that correspond to the spectral frequency and dissipation rate respectively (note that the mode lifetime is $\tau_{n} = 1/\Gamma_{n}$). These quantities, in turn, determine the mode quality factor via $Q = \omega_{n}/\Gamma_{n}$.

The QNM eigenfrequencies $\tilde{\omega}_n =\omega_n- \mathrm{i} \gamma_n$ are generally complex in character by virtue of losses
%~\cite{sauvan2013theory}
\footnote{Notably however, static/longitudinal modes with $\tilde{\omega}_\mu = 0 \in \mathbb{R}$ can occur, and the significance of their non-resonant contribution in QNM theories has been investigated in the literature~\cite{sauvan2021quasinormal, besbes2022role}.}.
%\textcolor{red}{Can static modes be removed by adding a small loss to the background medium? Are they correctly recovered in the case of dispersive PMLs?}
Due to their exponential decay in time ($\gamma_n < 0$), the QNMs exponentially diverge in space at $\vert\bm{r}\vert \rightarrow \infty$~\cite{sauvan2022norm}. Moreover, since they satisfy an outgoing-wave condition, the QNMs can only be used to expand scattered fields. It is therefore natural to adopt a background-field formulation which splits the total field into $\bm{E}(\bm{r},\omega) = \bm{E}_\mathrm{b}(\bm{r},\omega) + \bm{E}_\mathrm{sca}(\bm{r},\omega)$. Here, the background or incident field $\bm{E}_\mathrm{b}(\bm{r},\omega)$ satisfies Maxwell's equations in the background medium,
\begin{equation}
    \nabla \times \left( \mu^{-1}_\mathrm{b} \nabla \times \bm{E}_\mathrm{b}\right) - \frac{\omega^2}{c^2} \varepsilon_\mathrm{b}\bm{E}_\mathrm{b} = \mathrm{i}\omega\mu_0 \bm{j}_\mathrm{b} \ ,
\end{equation}
and could be a solution of the homogeneous equation or generated by some current density $\bm{j}_\mathrm{b}$ outside the resonator. Note that in the case of a non-trivial background geometry, scattering effects are already accounted for in the background part of the field, as sketched in Fig.~\ref{Fig_PML}(a). The scattered field $\bm{E}_\mathrm{sca}(\bm{r},\omega)$ then satisfies Maxwell's equations in the full system, see Fig.~\ref{Fig_PML}(b), with an effective current density,
\begin{align}
    \nabla \times &\left( \mu^{-1} \nabla \times \bm{E}_\mathrm{sca}\right) - \frac{\omega^2}{c^2} \varepsilon\bm{E}_\mathrm{sca} \notag \\
    &= \frac{\omega^2}{c^2} \Delta \varepsilon \bm{E}_\mathrm{b} - \nabla \times \left( \Delta \mu^{-1} \nabla \times \bm{E}_\mathrm{b}\right)\ , 
\end{align}
where $\Delta \varepsilon = \varepsilon - \varepsilon_\mathrm{b}$, $\Delta \mu^{-1} = \mu^{-1} - \mu^{-1}_\mathrm{b}$ [$\mathrm{supp}(\Delta \varepsilon(\cdot,\omega )/\Delta \mu^{-1}(\cdot,\omega) ) \subseteq \Omega_\mathrm{res}$], and the same outgoing-wave condition as for the QNMs. 

With regard to the practical calculation of QNMs and the formulation of a modal theory, two problems arise that need to be addressed. Considering their calculation, an outgoing-wave condition is usually imposed as an asymptotic one for the fields at $\vert\bm{r}\vert \rightarrow \infty$, while in numerical calculations the domain must be truncated. Using the QNMs in an expansion of the scattered field $\bm{E}_\mathrm{sca}$ at real frequencies $\omega$, it is apparent that this expansion cannot be valid at positions far away from the resonator due to their spatial divergence. In fact, it is known that the QNMs can only be used to expand the scattered field and the system Green's function for positions inside the resonator, where they appear to form a complete set, provided that the resonator is defined by a discontinuity in the refractive index in space~\cite{leung1994completeness,leung1996dielectric,lee1999dyadic}. However, even this is only true for trivial homogeneous backgrounds, and becomes generally invalid for more complex background geometries, such as the one sketched in Fig.~\ref{Fig_PML}, due to branch cuts of the Green's function~\cite{yan2018rigorous,sauvan2022norm} (meaning that in such cases, the QNMs themselves generally do not form a complete set inside the resonator). A general QNM expansion of the scattered field $\bm{E}_\mathrm{sca}(\bm{r},\omega)$ would thus have the form
\begin{align}\label{eqn:QNM_expansion_general}
    \bm{E}_\mathrm{sca}(\bm{r},\omega) = \sum_n \alpha_n(\omega) \bm{E}_n(\bm{r}) + \bm{E}_\mathrm{nr}(\bm{r},\omega) \, , 
\end{align}
in which $\bm{E}_\mathrm{nr}(\bm{r},\omega)$ is a non-resonant continuum contribution that prevents the formulation of a discrete QNM modal theory. Both challenges, the truncation of the domain and the incompleteness of the QNMs, can be addressed by using exterior complex coordinate transformations, or equivalently, PMLs. 
% To perform quantization, we require QNMs of Eq.~\eqref{classicalQNM} to form a complete set and normalized, which is possible through proper mode regularization. To this aim, we consider a smooth-shaped nanoplasmonic resonator~(as in Fig.~\ref{Fig_PML}) subjected to a PML layer, whose modes are QNMs that satisfy \eqref{classicalQNM}.

PMLs were originally introduced in computational electrodynamics in order to simulate open boundaries~\cite{berenger1994perfectly}. They are defined as material layers that surround a finite domain such that the dynamics therein are unaffected, \textit{i.e.}, outgoing waves are transmitted, without reflection, into the PML region $\Omega_\mathrm{PML}$ and are exponentially damped within them. The outgoing wave condition is thus effectively transformed into a Dirichlet (\textit{i.e.}, perfect electrical conductor) boundary condition at $\vert\bm{r}\vert \rightarrow \infty$. Due to the fast exponential damping of the waves, it is a good approximation to consider PMLs with finite thickness only, so that the infinite space $\mathbb{R}^3$ has been effectively mapped to a finite domain $\Omega$, in which we solve the scattered-field problem. This situation is sketched in Fig.~\ref{Fig_PML}(c). We denote the transformed material parameters of the mapped system by $(\tilde{\varepsilon}(\bm{r},\omega), \tilde{\mu}(\bm{r},\omega))$, whereas the inner domain $\Omega \setminus \Omega_\mathrm{PML}$ is not affected by this transformation. When using PMLs, however, there are some limitations. In particular, the material parameters are required to be analytic functions in the directions perpendicular to the PML boundaries (connected to the idea of PMLs as complex coordinate transformations), implying that the PMLs must be chosen in such a manner that waveguides enter with an angle of $90^\circ$, and periodic photonic crystal environments cannot be treated without further considerations~\cite{oskooi2008failure}. In this case, it is believed that the QNMs are not complete inside the resonator~\cite{sauvan2022norm}, but notably, efforts to adapt QNM methodology for coupled cavity-waveguide systems have been made via the introduction of a non-local boundary condition~\cite{kristensen2014calculation}.

We briefly review the formulation of PMLs in terms of complex coordinate transformations, focusing specifically on the case of Cartesian PMLs as sketched in Fig.~\ref{Fig_PML}(c), however, generalization is straightforward. %The inner wall at $r=R_\text{in}$ is soft and capable of transmitting electromagnetic fields into the absorbing layer, while to the outer wall at $r = R_\text{out}$ we associate a perfect electrical conductor boundary condition, $\bm{n}\times\tilde{\bm{E}}_l(R_\text{out})=0$, where $\bm{n}$ is a unit vector perpendicular to the spherical surface.
For each coordinate $x_i$, the absorbing PML region is itself characterized by a (generally) coordinate-dependent absorption function $\sigma_i(x_i) \geq 0$ with $\mathrm{supp}(\sigma_i) \subseteq \Omega_\mathrm{PML}$. 
%subject to boundary condition $\bm{n}\times\tilde{\bm{E}}_l(R_\text{out})=0$; $\bm{n}$ being the unit vector perpendicular to the PML surface for $r=R_\text{out}$.
%Based on this absorption function, and
%Assuming homogeneity in the azimuthal and polar angular coordinates, we set
% \begin{equation*}
%     f_i(x_i)=\int^{x_i}\mathrm{d}x'_i\, \sigma_i(x'_i)\, ,
% \end{equation*} 
We can define the complex coordinate transformation corresponding to the PMLs as
\begin{equation}
    x_i \mapsto \tilde{x}_i(x_i)= x_i + \frac{\mathrm{i}}{\omega} \int^{x_i}\mathrm{d}x'_i\, \sigma_i(x'_i)\, ,
    \label{Eq:PML_Absorption}
\end{equation}
under which any outgoing wave with wavevector component $k_i$ becomes exponentially damped in the direction of $x_i$, 
\begin{align}
    \exp\{\mathrm{i} k_i x_i\} \mapsto &\exp\{\mathrm{i} k_i \tilde{x}_i(x_i)\} \notag \\
    &= \mathrm{e}^{\mathrm{i} k_i x_i- \frac{1}{c} \cos(\theta_i) \int^{x_i}\mathrm{d}x'_i\, \sigma_i(x'_i)} \, ,
\end{align}
with $\cos(\theta_i) = c k_i/\omega$.
PMLs were initially introduced to damp waves with real frequencies $\omega$, but they can also be used to modify the behaviour of complex-frequency waves with $\tilde{\omega} = \omega - \mathrm{i}\gamma$, \textit {viz.} 
\begin{equation}
    |\exp\{\mathrm{i} \tilde{k}_i \tilde{x}_i(x_i)\}| = \mathrm{e}^{- \frac{1}{c} \cos(\theta_i) \left(\int^{x_i}\mathrm{d}x'_i\, \sigma_i(x'_i) - \gamma x_i \right)} \, ,
\end{equation}
and can therefore be utilized for the computation of QNMs.
Note that the wave is only damped if the expression in the brackets on the right-hand side is positive, \textit{i.e.}, $\sigma_i(x_i)$ has to be chosen large enough. In theory, PMLs can be made arbitrarily attenuating by merely increasing $\sigma_i(x_i)$ and its thickness. As numerical calculations usually involve a discretization of space (via finite differences or finite elements), spurious numerical reflections occur, and one typically adopts a gradually increasing absorption profile $\sigma_i(x_i)$ to minimize them.   
By applying this transformation to all coordinates $x_i$, we obtain a coordinate transformation that damps all outgoing waves and has the Jacobian
\begin{equation}
    J_{i j}=\frac{\partial\tilde{x}_i}{\partial x_j} = \left( 1 + \frac{\mathrm{i}}{\omega} \sigma_i(x_i) \right) \delta_{i j} \ .
    \label{Eq:PML_Jacobian}
\end{equation}
By using the correspondence between coordinate transformations and effective materials in electrodynamics~\cite{ward1996refraction}, Maxwell's equations with the material parameters $(\varepsilon,\mu)$ in the coordinates $\tilde{x}_i$ can be rewritten in the real coordinates $x_i$ with effective materials $(\tilde{\varepsilon}, \tilde{\mu})$ using the transformations
\begin{align}\label{eqn:PML_parameter_transformation}
    \tilde{\varepsilon}= \det(J) J^{-1} \varepsilon J^{-\mathrm{T}} \ , \ \tilde{\mu} = \det(J) J^{-1} \mu J^{-\mathrm{T}} \ ,
\end{align}
where $\det(J)$ is the determinant of the Jacobian. The transformed QNM eigenvalue problem, corresponding to Fig.~\ref{Fig_PML}(c), is thus given by
\begin{align}\label{eqn:classicalQNM_transformed}
    \begin{bmatrix}
    0 & - \mathrm{i} \mu_0^{-1}\tilde{\mu}^{-1}(\bm{r},\tilde{\omega}_n)\nabla\times \\
    \mathrm{i} \varepsilon_0^{-1}\tilde{\varepsilon}^{-1}(\bm{r},{\tilde \omega}_{n})\nabla\times & 0 \\
    \end{bmatrix}
    \begin{bmatrix}
    \tilde{\bm{H}}_n \\
    \tilde{\bm{E}}_n
    \end{bmatrix} \notag \\
    = {\tilde \omega}_{n}
    \begin{bmatrix}
    \tilde{\bm{H}}_n \\
    \tilde{\bm{E}}_n
    \end{bmatrix} .
\end{align}
The solutions of this transformed and PML-truncated eigenvalue problem are a discrete set of modes $\tilde{\bm{E}}_n(\bm{r})$ that can be grouped into QNM-like modes $\tilde{\bm{E}}_n^\mathrm{QNM}(\bm{r})$, which are good approximations of the true QNMs of the system and do not depend on the parameters of the PMLs, and so-called numerical or PML modes $\tilde{\bm{E}}_n^\mathrm{PML}(\bm{r})$ that depend sensitively on the PML parameters and which originate from the continuous non-resonant contribution in Eq.~\eqref{eqn:QNM_expansion_general}. Hence, by introducing finite-thickness PMLs, this background is effectively mapped to a discrete set of modes~\cite{olyslager2004discretization}. Since the operator in Eq.~\eqref{eqn:classicalQNM_transformed} is not hermitian, the completeness of these eigenmodes is not always guaranteed and we must further assume the absence of exceptional points where two eigenvectors coalesce. In this case, the QNMs and PML modes form a complete set, even in the presence of non-trivial backgrounds, and the PML-regularized scattered field $\tilde{\bm{E}}_\mathrm{sca}$ can be expanded in this discrete set of modes,
\begin{align}\label{eqn:QNM_expansion_PML}
    \tilde{\bm{E}}_\mathrm{sca}(\bm{r},\omega) &= \sum_n \alpha_n^\mathrm{QNM}(\omega)\tilde{\bm{E}}_n^\mathrm{QNM}(\bm{r}) + \alpha_n^\mathrm{PML}(\omega)\tilde{\bm{E}}_n^\mathrm{PML}(\bm{r}) \notag \\
    &= \sum_n \alpha_n(\omega)\tilde{\bm{E}}_n(\bm{r}) \, , 
\end{align}
where, henceforth, we shall not distinguish between the QNM-like and PML modes. Mathematically, they arise as a common set of solutions to the eigenvalue problem in Eq.~\eqref{eqn:classicalQNM_transformed}, and should be treated on an equal footing for the purpose of further analysis.
%\textcolor{red}{Studying if expansion can still be used at an EP might be interesting? \\ DDAC: This is a good suggestion. Whilst I would avoid elaborating on the issue in this manuscript, it is certainly relevant to, e.g., studies of exceptional points in active metamaterials, where they can arise at frequencies where loss is exactly compensated for by gain. Keep this in mind for future discussions.}

In the context of non-dispersive magnetic permeabilities $\mu = \mu(\bm{r})$, Yan {\it et al.}~\cite{yan2018rigorous} have developed an approach for rigorous QNM analysis of lossy resonators, relying on an auxiliary-field eigenvalue technique with PML regularization, within the framework of a finite-element analysis that enables efficient numerical calculations.
Their scheme effectively linearizes the non-linear eigenvalue problem embodied by Eq.~\eqref{eqn:classicalQNM_transformed} for dispersive materials, described using an $N$-pole, Drude-Lorentz model:
\begin{equation}\label{eqn:Drude_Lorentz}
    \varepsilon(\bm{r},\omega) = \varepsilon_\infty(\bm{r}) -\varepsilon_\infty(\bm{r})\sum_{l=1}^N \frac{\omega_{\text{p},l}^2(\bm{r})}{\omega^2-\omega_{0,l}^2(\bm{r})+ \mathrm{i} \gamma_l(\bm{r}) \omega} ,
\end{equation}
in which $\varepsilon_\infty(\bm{r})$ is the permittivity in the high-frequency limit, while $\gamma_l(\bm{r})$, $\omega_{\text{p},l}(\bm{r})$ and $\omega_{0,l}(\bm{r})$ are damping rates, plasma and resonance frequencies pertaining to the $l$th pole, respectively. The method has been comprehensively benchmarked and compared to other approaches in Ref.~\cite{lalanne2019quasinormal}. Whilst the quantization scheme that we will discuss in \S~\ref{Sec:Quantum_QNMs} can in principle be used for arbitrary dispersions of $(\tilde{\varepsilon},\tilde{\mu})$, the approach of Yan {\it et al.} facilitates practical calculations in the important special cases of dielectric and plasmonic cQED, and formulae obtained within it can be used in the quantized theory for these cases as well (such as, for instance, the analytic excitation coefficients of the modes for an external driving field $\bm{E}_\mathrm{b}$). The modes, both QNMs and PML modes, can be normalized via the condition
\begin{align}
    \int_{\Omega}\mathrm{d}^3 r\, \left[\partial_\omega[ \omega \varepsilon_0\tilde{\varepsilon}(\bm{r},\omega)]_{\omega=\tilde{\omega}_n}\tilde{\bm{E}}_n\cdot\tilde{\bm{E}}_{n}-\mu_0\tilde{\mu}(\bm{r})\tilde{\bm{H}}_n\cdot\tilde{\bm{H}}_{n}\right] \notag \\
    = 1 \, ,
\end{align}
in which the volume integration is performed over the entire computational domain $\Omega$, including the PML regions $\Omega_{\mathrm{PML}}$, yet the normalization of the QNMs does not depend on the PML parameters~\cite{sauvan2022norm}. At this point, it is worth noting that different normalization procedures have been proposed in the literature, whose validity and equivalence have been the subject of intense debate~\cite{kristensen2015normalization,muljarov2017comment,kristensen2017reply,lalanne2020mode}; for a mathematical and numerical comparison, we refer the interested reader to Ref.~\cite{sauvan2022norm}. For the excitation coefficients in Eq.~\eqref{eqn:QNM_expansion_PML}, an analytical form can be derived~\cite{yan2018rigorous}:
\begin{align}
    \alpha_n(\omega) = \frac{\tilde{\omega}_n}{\tilde{\omega}_n - \omega} \varepsilon_0&\langle \tilde{\bm{E}}^*_n | \Delta\varepsilon(\bm{r},\tilde{\omega}_n) | \bm{E}_\mathrm{b} \rangle_{L^2(\Omega_\mathrm{res})} \notag \\
    + \varepsilon_0&\langle  \tilde{\bm{E}}^*_n | \varepsilon_\mathrm{b}(\bm{r},\omega) - \varepsilon_\infty(\bm{r)} | \bm{E}_\mathrm{b} \rangle_{L^2(\Omega_\mathrm{res})} \notag \\
    -\frac{\tilde{\omega}_n}{\tilde{\omega}_n - \omega} \mu_0&\langle \tilde{\bm{H}}^*_n | \Delta\mu(\bm{r}) | \bm{H}_\mathrm{b} \rangle_{L^2(\Omega_\mathrm{res})} \, .
\end{align}
A modal expansion as in Eq.~\eqref{eqn:QNM_expansion_PML} is also possible in the time domain, with coefficients $\xi_n(t)$ related to $\alpha_n(\omega)$ via a Fourier transform and satisfying the equation
\begin{align}
    \frac{\mathrm{d}\xi_n}{\mathrm{d}t}= &-\mathrm{i}\tilde{\omega}_n \xi_n(t) + \mathrm{i}\tilde{\omega}_n \varepsilon_0\langle \tilde{\bm{E}}_n^* | \varepsilon(\tilde{\omega}_n) - \varepsilon_\mathrm{b} | \bm{E}_\mathrm{b}(t) \rangle \notag \\
    &\hspace{-0.5cm}- \varepsilon_0\langle \tilde{\bm{E}}_n^* | \varepsilon_\mathrm{b} - \varepsilon_\infty | \frac{\mathrm{d}}{\mathrm{d}t}\bm{E}_\mathrm{b}(t) \rangle - \mathrm{i}\tilde{\omega}_n \mu_0\langle \tilde{\bm{H}}_n^* | \Delta \mu | \bm{H}_\mathrm{b}(t) \rangle \notag \\
    =  &-\mathrm{i}\tilde{\omega}_n \xi_n(t) + D_n(t) \ . \label{eqn:QNM_coeffs_time_domain}
\end{align}
The system Green's function can also be expanded in the modes according to
\begin{align}\label{eqn:Greens_function_expansion}
    G(\bm{r},\bm{r}',\omega) = \sum_n A_n(\omega) \tilde{\bm{E}}_n(\bm{r}) \otimes \tilde{\bm{E}}_n(\bm{r}) \ , 
\end{align}
with $A_n = \varepsilon_0 c^2 / \omega(\tilde{\omega}_n-\omega)$ or $A_n = \varepsilon_0 c^2 / \tilde{\omega}_n(\tilde{\omega}_n-\omega)$, depending on the pole structure of the Green's function at $\omega = 0$~\cite{kristensen2020modeling}.
It is worth noting that although this auxiliary-field eigenvalue approach was originally introduced for systems with non-dispersive magnetic permeabilities $\mu(\bm{r})$, its generalization for media with strongly frequency-dependent permeabilities (as relevant for magnonic cQED) may be feasible via the introduction of suitable auxiliary fields for the magnetic response. In the following, we assume that such a complete set of suitably regularized and normalized modes can be ascertained, and explore their quantization for arbitrary, linear, 3D, magnetodielectric systems.

\section{\label{Sec:Quantum_QNMs} Quantization of the Regularized Quasinormal Modes}

We now introduce a quantization scheme for the regularized QNMs of arbitrary, 3D, linear, dissipative (with possible gain components) and dispersive nanoresonators, as defined in \S~\ref{Sec:Classic_QNMs}. In our approach, we will quantize the eigenmodes of the PML-truncated domain in Eqs.~\eqref{eqn:classicalQNM_transformed} and \eqref{eqn:QNM_expansion_PML}. As already mentioned, this set of eigenmodes includes QNM-like and PML modes, which following our earlier remarks, shall simply be referred to as modes without further differentiation.
%For the sake of simplicity, we will simply refer to them as modes from now on, as they are treated equally.

Similar to previous work~\cite{franke2019quantization, franke2020nonlinear, franke2022coupled}, our approach is predicated on the formalism of mQED, which offers a phenomenological Green's function quantization scheme for electromagnetic fields in the presence of spatially inhomogeneous, lossy and dispersive media~\cite{gruner1996mqed,dung1998mqed,scheel1998qed,dung2000mqed, raabe2007unified}, later generalized for systems with gain~\cite{matloob1997electromagnetic,raabe2008qed}. However, in order to facilitate a rigorous application of the theory, a brief discussion concerning some important prerequisites is warranted. The complex permittivity $\varepsilon(\bm{r},\omega) = \varepsilon_\mathrm{r}(\bm{r},\omega) + \mathrm{i} \varepsilon_\mathrm{i}(\bm{r},\omega)$ and inverse permeability $\mu^{-1}(\bm{r}, \omega) = \kappa(\bm{r},\omega) = \kappa_\mathrm{r}(\bm{r},\omega) + \mathrm{i} \kappa_\mathrm{i}(\bm{r},\omega)$ are required to satisfy the Kramers-Kronig relations (\textit{i.e.}, they are analytic in the closed, upper-half, complex $\omega$ plane and 
$\zeta(\bm{r},\omega) \rightarrow 1$ for $|\omega| \rightarrow\infty$ ($\zeta \in \{\varepsilon,\kappa \}$). This is not really a requirement of the quantum theory, but arises from the physical necessity of causality. Furthermore, they should be reciprocal, $\zeta(\bm{r},\omega) = \zeta^\mathrm{T}(\bm{r},\omega)$, and satisfy the reflection symmetry condition $\zeta_{i j}(\bm{r},-\omega^*) = \zeta^*_{i j}(\bm{r}, \omega)$, which guarantees their reality in the time domain. These properties ensure that the fundamental QED commutation relations between the electric and magnetic field operators are maintained. If the material parameters $\zeta(\bm{r},\omega)$ of the system are reciprocal, it readily follows from Eq.~\eqref{eqn:PML_parameter_transformation} that the PML-transformed parameters $\tilde{\zeta}(\bm{r},\omega)$ are as well. It should be noted that there have been generalizations of mQED for the cases of non-reciprocal materials and materials with magnetoelectric crossing effects~\cite{buhmann2012macroscopic}, which are, however, beyond the scope of this work.

When incorporating dispersive or non-dispersive PMLs in the quantization, two problems arise. For dispersive PMLs, it is easy to show from Eq.~\eqref{eqn:PML_parameter_transformation} and the Jacobian in Eq.~\eqref{Eq:PML_Jacobian} that the PML-transformed parameters $\tilde{\zeta}(\bm{r},\omega)$ satisfy reflection symmetry, which preserves the important property that modes with $\Re(\tilde{\omega}_n) \neq 0$ always appear in pairs $(\tilde{\bm{E}}_\mu,\tilde{\omega}_\mu)$, $(\tilde{\bm{E}}^*_n,-\tilde{\omega}^*_n)$ and that all the modes have a non-positive imaginary part $\Im(\tilde{\omega}_n) \leq 0$. Due to the $\omega$-dependence of the coordinate transformation, the material parameters have a pole at $\omega = 0$ and consequently do not satisfy the Kramers-Kronig relations (\textit{i.e.}, they do not correspond to a causal medium). To resolve this issue, we could construct causal PMLs~\cite{kuzuoglu2002frequency} by making the replacement $\omega \rightarrow \omega + \mathrm{i} \eta$ in Eq.~\eqref{Eq:PML_Absorption} with a small $\eta > 0$, thereby shifting the pole into the lower, complex $\omega$ half-plane. Note that these causal PMLs also work, in principle, in the static case $\omega = 0$, due to the finite values of the transformed material parameters at this point~\cite{kuzuoglu2002frequency}, whereas the usual PMLs fail and cannot be used to reproduce static modes~\cite{besbes2022role}. However, we are only interested in the fields in the inner, unmapped domain $\Omega \setminus \Omega_\mathrm{PML}$, and by using the properties of the Green's function we will show that the correct commutator is maintained within this region even if it is not in the PML region $\Omega_\mathrm{PML}$, so we expect that the use of PMLs as defined in Eq.~\eqref{Eq:PML_Absorption} should suffice in most cases where static modes can be neglected. When using non-dispersive PMLs, the PML-transformed parameters no longer satisfy reflection symmetry, and since they do not afford the damping of waves with non-positive frequencies, the QNMs with $\Re(\tilde{\omega}_n) \leq 0$ are not correctly recovered and PML modes with $\Im(\tilde{\omega}_n) > 0$ occur, which would grow exponentially in time~\cite{sauvan2022norm}. In practice, one is often interested in the spectral response of a system only within a finite frequency interval $\Delta \omega$,
%\in \mathbb{R}_+$.
in which case non-dispersive PMLs could be used provided that the contributions of the modes with $\Re(\tilde{\omega}_n) \leq 0$ are negligible in $\Delta \omega$. With the same argument, one can also justify the use of material parameters $\zeta(\bm{r},\omega)$ or PMLs that do not satisfy the Kramers-Kronig relations (\textit{e.g.}, non-dispersive and complex), provided that the true, physical parameters are sufficiently well approximated within $\Delta\omega$. 
%\textcolor{red}{Cite Franke thesis?} 

The formalism of mQED is based on the dyadic Green's function of the system, defined by 
\begin{align}\label{eqn:Greens_function}
    \nabla \times \tilde{\kappa} \nabla \times G(\bm{r},\bm{r}',\omega) - \frac{\omega^2}{c^2} \tilde{\varepsilon} G(\bm{r},\bm{r}',\omega) = \bm{I} \delta(\bm{r}-\bm{r}') \, .
\end{align}
If the system exhibits loss only, then the Green's function is analytic in the upper $\omega$ half-plane and has a set of discrete poles (corresponding to the QNMs), as well as possible branch cuts in the lower half-plane in the case of non-trivial background geometries. The introduction of gain components in the system shifts up the poles in the complex $\omega$ plane and a prerequisite to apply the theory for such active systems is that all the poles remain located in the lower half-plane. Again, this is not so much a fundamental limitation of the quantum theory but is, rather, a constraint on its range of validity for systems with linear amplification~\cite{raabe2008qed}. The Green's function thus also satisfies the Kramers-Kronig relations, and besides that, reflection symmetry and the reciprocity $G(\bm{r},\bm{r}',\omega) = G^\mathrm{T}(\bm{r}',\bm{r},\omega)$.

From Eq.~\eqref{eqn:Greens_function}, one can derive the useful relation
\begin{align}\label{eqn:Im_Greens}
    &\Im(G(\bm{r},\bm{r}',\omega)) =  \frac{\omega^2}{c^2} \int_\Omega \mathrm{d}^3 s\, G^\dagger(\bm{s},\bm{r},\omega) \tilde{\varepsilon}_\mathrm{i}(\bm{s},\omega) G(\bm{s},\bm{r}',\omega)  \notag\\
    &- \int_\Omega \mathrm{d}^3 s\, \left( \nabla \times G(\bm{s},\bm{r},\omega)\right)^\dagger\tilde{\kappa}_\mathrm{i}(\bm{s},\omega) \left( \nabla \times G(\bm{s},\bm{r}',\omega)\right) \, , 
\end{align}
in which the curl operator has to be applied column-wise, and the real and imaginary parts of tensorial fields are defined as their hermitian and anti-hermitian parts, respectively, which reduce to the component-wise real and imaginary parts due to reciprocity. Note that although the integrals on the right-hand side have to be taken over the whole space, including the PML domain $\Omega_\mathrm{PML}$, the left-hand side does not depend on the PML parameters as long as $\bm{r}, \bm{r}' \in \Omega \setminus\Omega_\mathrm{PML}$, given that the PMLs (ideally) do not affect the Green's function in the unmapped domain.

% The second ingredient that we need to formulate the modal quantum theory is an expansion of the Green's function in terms of the modes in Eq.~\eqref{eqn:classicalQNM_transformed}. By using a residue-decomposition approach, one can derive the expansion
% \begin{align}\label{eqn:Greens_function_expansion}
%     G(\bm{r},\bm{r}',\omega) = \sum_n A_n(\omega) \tilde{\bm{E}}_n(\bm{r}) \otimes \tilde{\bm{E}}_n(\bm{r}) \ , 
% \end{align}
% with $A_n = \varepsilon_0 c^2 / \omega(\tilde{\omega}_n-\omega)$ or $A_n = \varepsilon_0 c^2 / \tilde{\omega}_n(\tilde{\omega}_n-\omega)$, where the non-uniqueness of the expansion stems from the overcompleteness of the set of modes~\cite{lalanne2018light}. 

In mQED, the frequency components of the quantized electric field operator $\hat{\bm{E}}(\bm{r},\omega)$ satisfy the equation
\begin{align}
    \nabla \times \tilde{\kappa}\nabla \times \hat{\bm{E}}(\bm{r},\omega) - \frac{\omega^2}{c^2} \tilde{\varepsilon}\hat{\bm{E}}(\bm{r},\omega) = i \omega \mu_0 \hat{\bm{j}}_\mathrm{F}(\bm{r},\omega) \, . 
\end{align}
In the case of magnetodielectric materials with loss only, the noise/fluctuation current density $\hat{\bm{j}}_\mathrm{F}(\bm{r},\omega)$ introduced on the right-hand side, in accordance with the fluctuation-dissipation theorem, can be written as
\begin{align}\label{eqn:noise_current_loss}
    \hat{\bm{j}}_\mathrm{F}(\bm{r},\omega) = &\omega \sqrt{\frac{\hbar \varepsilon_0}{\pi} \tilde{\varepsilon}_{\mathrm{i},j}(\bm{r},\omega)} \hat{f}_{\mathrm{e},j}(\bm{r},\omega) \bm{e}_j  \\
    &+ \nabla \times \bigg( \sqrt{-\frac{\hbar}{\pi\mu_0} \tilde{\kappa}_{\mathrm{i},j}(\bm{r},\omega)} \hat{f}_{\mathrm{m},j}(\bm{r},\omega) \bm{e}_j   \bigg) \, , \notag
\end{align}
in which we have used the Einstein summation convention and $\hat{f}_{\lambda, j}(\bm{r},\omega)$ ($\hat{f}^\dagger_{\lambda, j}(\bm{r},\omega)$) are bosonic annihilation (creation) operators of the joint (polaritonic) excitations of the field-dressed medium. The latter satisfy the commutation relation\cite{dung1998mqed}
\begin{equation}
    \left[\hat{f}_{\lambda, i}(\bm{r},\omega), \hat{f}^\dagger_{\lambda', j}(\bm{r}',\omega') \right]= \delta_{\lambda \lambda'}\delta_{ij}\delta(\bm{r}-\bm{r}')\delta(\omega-\omega') \, ,
    \label{Eq:Bosonic_Commutation}
\end{equation}
and diagonalize the material-field Hamiltonian:
\begin{align}\label{eqn:mQED_Hamitonian}
    \hat{H}_\mathrm{EM} = \int_\Omega \mathrm{d}^3r \int_0^\infty\mathrm{d}\omega\, \hbar \omega \bigg( &\hat{f}^\dagger_{\mathrm{e}, j}(\bm{r},\omega)\hat{f}_{\mathrm{e}, j}(\bm{r},\omega) \\
    &+ \hat{f}^\dagger_{\mathrm{m}, j}(\bm{r},\omega)\hat{f}_{\mathrm{m}, j}(\bm{r},\omega)\bigg) \, . \notag
\end{align}
Note that in Eq.~\eqref{eqn:noise_current_loss}, we have assumed scalar material parameters $\zeta(\bm{r},\omega)$ and Cartesian PMLs, for which the PML-transformed parameters are diagonal with $\tilde{\zeta}_{\mathrm{i}, j j} = \tilde{\zeta}_{\mathrm{i},j}$. However, the analysis can easily be generalized for the non-diagonal case by replacing $\sqrt{\tilde{\zeta}_{\mathrm{i}, j}(\bm{r},\omega)}$ and the standard basis vectors $\bm{e}_j$ by the eigenvalues and eigenvectors of the matrices $\Im(\zeta(\bm{r},\omega))^{\frac{1}{2}}$, respectively (\textit{i.e.}, by transforming to a system of principal axes).
Explicitly, this would mean replacing Eq.~\eqref{eqn:noise_current_loss} by
\begin{align}
    &\hat{\bm{j}}_\mathrm{F}(\bm{r},\omega) = \omega\sqrt{\frac{\hbar\varepsilon_0}{\pi}} (\Im(\tilde{\varepsilon}(\bm{r},\omega))^\frac{1}{2} \cdot \hat{\bm{f}}_\mathrm{e}(\bm{r},\omega) \notag \\
    &= \omega\sqrt{\frac{\hbar\varepsilon_0}{\pi}}(\Im(\tilde{\varepsilon}(\bm{r},\omega))^\frac{1}{2}_i \bm{e}_i(\bm{r},\omega)\left(\bm{e}^*_i(\bm{r},\omega) \cdot \hat{\bm{f}}_\mathrm{e}(\bm{r},\omega)\right) \notag \\
    &= \omega\sqrt{\frac{\hbar\varepsilon_0}{\pi}}(\Im(\tilde{\varepsilon}(\bm{r},\omega))^\frac{1}{2}_i \bm{e}_i(\bm{r},\omega) \hat{\tilde{f}}_\mathrm{e}(\bm{r},\omega) \notag \ , 
\end{align}
in which $(\Im(\tilde{\varepsilon}(\bm{r},\omega))^\frac{1}{2}_i$ are the eigenvalues of the matrix $\Im(\tilde{\varepsilon}(\bm{r},\omega)^\frac{1}{2}$ and $\bm{e}_i(\bm{r},\omega)$ its eigenvectors, which are complete and orthonormal since $\Im(\tilde{\varepsilon}(\bm{r},\omega))^\frac{1}{2}$ is a hermitian matrix. Note that the eigenvectors are the same as those of the matrix $\Im(\tilde{\varepsilon}(\bm{r},\omega))$. The transformation would work in the same manner for the magnetic part, although the eigenvectors could be different. The Hamiltonian in Eq.~\eqref{eqn:mQED_Hamitonian} would have the same form in the transformed bosonic operators $\hat{\tilde{f}}_\mathrm{e/m}(\bm{r},\omega)$, as the scalar products are invariant under a change of basis.

The noise current defined in Eq.~\eqref{eqn:noise_current_loss} stands in accordance with the fluctuation-dissipation theorem and ensures preservation of the fundamental commutation relations of the electric and magnetic fields. These formulae can also be derived from the canonical quantization of a microscopic model in which the electric and magnetic material responses are described using separate oscillator fields~\cite{philbin2010canonical}. From this perspective, it is clear that the Fock space corresponding to these bosonic variables describes not only the electromagnetic degrees of freedom, but also those of the material excitations. 

In the case of materials exhibiting gain, the creation and annihilation operators interchange roles in the respective (finite) spatial and spectral regions. Eq.~\eqref{eqn:noise_current_loss} must then be replaced by
\begin{align}\label{eqn:noise_current_gain}
    \hat{\bm{j}}_\mathrm{F}(\bm{r},\omega) = \omega \sqrt{\frac{\hbar\varepsilon_0}{\pi}|\tilde{\varepsilon}_{\mathrm{i},j}(\bm{r},\omega) |} \bm{e}_j \bigg[ \Theta(\tilde{\varepsilon}_{\mathrm{i},j}(\bm{r},\omega)) \hat{f}_{\mathrm{e},j}(\bm{r},\omega) \notag \\ 
      + \Theta(-\tilde{\varepsilon}_{\mathrm{i},j}(\bm{r},\omega)) \hat{f}^\dagger_{\mathrm{e},j}(\bm{r},\omega) \bigg] \notag \\
      +\nabla \times \bigg( \sqrt{\frac{\hbar}{\pi\mu_0} | \tilde{\kappa}_{\mathrm{i},j}(\bm{r},\omega)|} \bm{e}_j \bigg[\Theta(-\tilde{\kappa}_{\mathrm{i},j}(\bm{r},\omega)) \hat{f}_{\mathrm{m},j}(\bm{r},\omega) \notag \\
      + \Theta(\tilde{\kappa}_{\mathrm{i},j}(\bm{r},\omega)) \hat{f}^\dagger_{\mathrm{m},j}(\bm{r},\omega) \bigg]   \bigg) \, ,
\end{align}
and the (normal-ordered) Hamiltonian in Eq.~\eqref{eqn:mQED_Hamitonian} by
\begin{align}\label{eqn:mQED_Hamiltonian_gain}
    \hat{H}_\mathrm{EM} = \int_\Omega \mathrm{d}^3r &\int_0^\infty\mathrm{d}\omega\, \hbar \omega  \\
    \cdot\bigg(&\mathrm{sgn}( \tilde{\varepsilon}_{\mathrm{i},j}(\bm{r},\omega)) \hat{f}_{\mathrm{e},j}^\dagger(\bm{r},\omega)\hat{f}_{\mathrm{e},j}(\bm{r},\omega) \notag \\
    + &\mathrm{sgn}(-\tilde{\kappa}_{\mathrm{i},j}(\bm{r},\omega)) \hat{f}_{\mathrm{m},j}^\dagger(\bm{r},\omega)\hat{f}_{\mathrm{m},j}(\bm{r},\omega) \bigg) \, . \notag
\end{align}
Note that excitations in the gain regions have a negative-energy contribution to the Hamiltonian. These expressions can again be derived from a microscopic model using inverted oscillator fields in the gain regions~\cite{amooghorban2011casimir}. However, it is worth underlining that the expressions in Eqs.~\eqref{eqn:noise_current_loss} and \eqref{eqn:noise_current_gain} cannot be derived from a generalized phenomenological quantization scheme that allows for general non-local material responses described by a conductivity tensor, whenever the permittivity and permeability are spatially inhomogeneous~\cite{raabe2008qed, raabe2007unified}. Since these expressions can be motivated from a microscopic model and provide analytic results for arbitrary magnetodielectric media, we will make use of them in this paper. 

By using the definition of the Green's function in Eq.~\eqref{eqn:Greens_function}, the total electric field operator in the Schr{\"o}dinger picture can now be written as
\begin{align}
    \hat{\bm{E}}(\bm{r},\omega) &= i \mu_0\int_0^\infty\mathrm{d}\omega\, \omega\int_\Omega \mathrm{d}^3 s\, G(\bm{r},\bm{s},\omega) \bm{j}_\mathrm{F}(\bm{s},\omega) + \mathrm{h.c.} \notag \\
    &= \hat{\bm{E}}^{(+)}(\bm{r}) + \hat{\bm{E}}^{(-)}(\bm{r}) \ ,
\end{align}
where $\hat{\bm{E}}^{(\pm)}(\bm{r})$ are the positive/negative frequency parts. By using Eq.~\eqref{eqn:noise_current_gain} and the identity in Eq.~\eqref{eqn:Im_Greens}, we can ascertain the commutator of the electric field operator:
\begin{align}\label{eqn:E_field_commutator}
  &\left[ \hat{E}^{(+)}_i(\bm{r}), \hat{E}^{(-)}_j(\bm{r}') \right] \notag \\
  &= \frac{\hbar \mu_0}{\pi} \int_0^\infty \mathrm{d}\omega\, \omega^2 \int_{\Omega}\mathrm{d}^3s\notag \\
  &\hspace{1cm}\cdot\bigg[ \frac{\omega^2}{c^2} G^\dagger_{i m}(\bm{s},\bm{r},\omega) \tilde{\varepsilon}_{\mathrm{i},m}(\bm{s},\omega) G_{m j}(\bm{s},\bm{r}',\omega) \notag \\
  &- (\nabla \times G(\bm{s},\bm{r},\omega))^\dagger_{i m} \tilde{\kappa}_{\mathrm{i},m}(\bm{s},\omega)(\nabla \times G(\bm{s},\bm{r'},\omega))_{m j}   \bigg]   \notag \\
  &= \frac{\hbar \mu_0}{\pi} \int_0^\infty \mathrm{d}\omega\, \omega^2 \Im(G(\bm{r},\bm{r}',\omega))_{i j}  \, ,
\end{align}
where the latter equation shows that the commutator is not affected by the PMLs for $\bm{r},\bm{r}' \in \Omega\setminus\Omega_{\mathrm{PML}}$. Note that the identity in Eq.~\eqref{eqn:Im_Greens} can only be used if the spatial integration is performed over the entire space $\Omega$ including the PMLs, so that the presence of the PMLs must necessarily be taken into account in the quantization.

By invoking the first equality and the Green's function expansion in Eq.~\eqref{eqn:Greens_function_expansion}, we can expand the commutator in terms of the modes:
\begin{align}\label{eqn:E_field_commutator_QNM}
    \left[ \hat{E}^{(+)}_i(\bm{r}), \hat{E}^{(-)}_j(\bm{r}') \right] = \sum_{n,n'} P_{n n'} \tilde{E}_{n,i}(\bm{r}) \tilde{E}^*_{n',j}(\bm{r}') \, ,
\end{align}
with 
\begin{align}
    P_{n n'} = \frac{\hbar \mu_0}{\pi} &\int_0^\infty \mathrm{d}\omega\, \omega^2\int_{\Omega}\mathrm{d}^3 s\, A_n(\omega) A_{n'}^*(\omega)  \\ 
    \cdot\bigg[ &\frac{\omega^2}{c^2} \tilde{\bm{E}}_n(\bm{s}) \tilde{\varepsilon_{\mathrm{i}}}(\bm{s},\omega)\tilde{\bm{E}}^*_{n'}(\bm{s})  \notag\\
    &-(\nabla \times\tilde{\bm{E}}_n(\bm{s})) \tilde{\kappa}_{\mathrm{i}}(\bm{s},\omega)(\nabla \times\tilde{\bm{E}}^*_{n'}(\bm{s})) \bigg] \, . \notag
\end{align}
These integrals describe the losses of the modes. Ultimately, our aim is to develop a suitable modal expansion of the electric field operator. As we consider general, anisotropic, magnetodielectric systems which could exhibit electric and magnetic losses in distinct spatial/frequency regions, we start by decomposing the QNM operators in electric/magnetic and loss/gain components; we thus introduce the four operators $(\hat{\xi}_{n,\mathrm{e},\mathrm{L}}, \hat{\xi}_{n,\mathrm{e},\mathrm{G}}, \hat{\xi}_{n,\mathrm{m},\mathrm{L}}, \hat{\xi}_{n,\mathrm{m},\mathrm{G}})$ for each mode and expand the electric field operator in the form 
\begin{align}
    \hat{\bm{E}}(\bm{r}) &= \sum_n \tilde{E}_{n}(\bm{r}) (\hat{\xi}_{n,\mathrm{e},\mathrm{L}} + \hat{\xi}^\dagger_{n,\mathrm{e},\mathrm{G}} + \hat{\xi}_{n,\mathrm{m},\mathrm{L}} + \hat{\xi}_{n,\mathrm{m},\mathrm{G}}^\dagger) \notag \\
    &\hspace{6cm}+ \mathrm{h.c.} \notag \\
    &= \hat{\bm{E}}^{(+)}(\bm{r}) + \hat{\bm{E}}^{(-)}(\bm{r}) \, ,
\end{align}
in which the modal operators can be expressed through the continuous bosonic operators:
\begin{align}\label{eqn:QNM_operataors_unsymmetrized_e}
    \hat{\xi}^{(\dagger)}_{n,\mathrm{e},\mathrm{L/G}} = \mathrm{i} \mu_0 \int_0^\infty\mathrm{d}\omega\, \omega^2 A_n(\omega)\int_{\Omega}\mathrm{d}^3 r\, \theta(\pm\tilde{\varepsilon}_{\mathrm{i},j}(\bm{r},\omega)) \notag\\ \sqrt{\frac{\hbar\varepsilon_0}{\pi}|\tilde{\varepsilon}_{\mathrm{i},j}(\bm{r},\omega) |} \bm{e}_j\cdot \tilde{\bm{E}}_n(\bm{r}) \hat{f}_{\mathrm{e},j}^{(\dagger)}(\bm{r},\omega) \, , \\
    \hat{\xi}^{(\dagger)}_{n,\mathrm{m},\mathrm{L/G}} = \mathrm{i} \mu_0 \int_0^\infty\mathrm{d}\omega\, \omega A_n(\omega)\int_{\Omega}\mathrm{d}^3 r\, \theta(\mp\tilde{\kappa}_{\mathrm{i},j}(\bm{r},\omega)) \notag\\ \sqrt{\frac{\hbar}{\pi \mu_0}|\tilde{\kappa}_{\mathrm{i},j}(\bm{r},\omega) |} \bm{e}_j\cdot (\nabla\times\tilde{\bm{E}}_n(\bm{r})) \hat{f}_{\mathrm{m},j}^{(\dagger)}(\bm{r},\omega) \, . \label{eqn:QNM_operataors_unsymmetrized_m}
\end{align}
Note that the hermitian conjugates of the gain operators $\hat{\xi}_{n,\mathrm{e/m},\mathrm{G}}$ have to be used in the expansion of the positive-frequency part of the electric field, in order to reproduce the correct commutation relation as given in Eq.~\eqref{eqn:E_field_commutator_QNM}. Although they are hermitian conjugates of modal operators, they still behave like superpositions of positive-frequency excitations ($\propto \mathrm{e}^{-\mathrm{i}\omega t}$), as can be seen from Eqs.~\eqref{eqn:QNM_operataors_unsymmetrized_e} and \eqref{eqn:QNM_operataors_unsymmetrized_m}, and the Heisenberg equations for the continuum operators $\hat{f}_{\mathrm{e/m},j}(\bm{r},\omega)$ in the gain regions with the Hamiltonian of Eq.~\eqref{eqn:mQED_Hamiltonian_gain}. This is analogous to the approach in Ref.~\cite{franke2022coupled}. In our case, however, we have included the PMLs into the quantization, which always exhibit gain. In the presence of gain materials, the vacuum correlation function of the electric field would be given by
\begin{align}\label{eqn:vacuum_correlation_PMLs}
    &\langle \hat{E}_i(\bm{r}) \hat{E}_j(\bm{r}') \rangle \\
    &= \frac{\hbar \mu_0}{\pi} \int_0^\infty \mathrm{d}\omega\, \omega^2 \int_{\Omega}\mathrm{d}^3 s\notag \\
    &\hspace{1cm}\cdot \bigg[ \frac{\omega^2}{c^2} G^\dagger_{i m}(\bm{s},\bm{r},\omega) |\tilde{\varepsilon}_{\mathrm{i},m}(\bm{s},\omega)| G_{m j}(\bm{s},\bm{r}',\omega) \notag \\
  &+ (\nabla \times G(\bm{s},\bm{r},\omega))^\dagger_{i m} |\tilde{\kappa}_{\mathrm{i},m}(\bm{s},\omega)|(\nabla \times G(\bm{s},\bm{r'},\omega))_{m j}   \bigg] \notag\ .  
\end{align}
Due to the moduli of the material parameters in this formula, the identity in Eq.~\eqref{eqn:Im_Greens} cannot be used to express the correlation function in terms of the imaginary part of the Green's function, which must be taken into account when, for example, calculating QE decay rates~\cite{franke2021fermi} or Casimir(-Polder) forces~\cite{sambale2010casimir}. At this point, it is also important to be aware of the fact that PMLs in 3D always possess gain components (\textit{i.e.}, not all of the components of the effective material parameters $\tilde{\varepsilon}, \tilde{\mu}$ have a positive imaginary part). In our approach, the presence of these gain components in the PMLs would impact the correlation function due to induced fluctuations and thus the quantum dynamics, as they would, for example, affect the spontaneous decay of a QE. To see this, we can write the vacuum correlation function in our PML-transformed system as $\langle \hat{E}_i(\bm{r}) \hat{E}_j(\bm{r}') \rangle = (\hbar\mu_0/\pi) \int\mathrm{d}\omega\,\omega^2\mathrm{Im}(G_{ij}(\bm{r},\bm{r}',\omega)) + C_\mathrm{gain}^\mathrm{res}(\bm{r},\bm{r}') + C_\mathrm{gain}^\mathrm{PML}(\bm{r},\bm{r}')$, with
\begin{align}
    C_\mathrm{gain}^\mathrm{res/PML}(\bm{r},\bm{r}') =\frac{2\hbar\mu_0}{\pi} \int_0^\infty\mathrm{d}\omega\, \omega^2 
    \int_{\Omega{\mathrm{res/PML}}} \mathrm{d}^3 s\, \notag \\ \theta(-\tilde{\varepsilon}_{\mathrm{i},m}(\bm{s},\omega))  \frac{\omega^2}{c^2} G^\dagger_{i m}(\bm{s},\bm{r},\omega) |\tilde{\varepsilon}_{\mathrm{i},m}(\bm{s},\omega)| G_{m j}(\bm{s},\bm{r}',\omega) \notag \\
    + \theta(\tilde{\kappa}_{\mathrm{i},m}(\bm{s},\omega))  (\nabla \times G(\bm{s},\bm{r},\omega))^\dagger_{i m} |\tilde{\kappa}_{\mathrm{i},m}(\bm{s},\omega)| \notag \\ 
    \cdot(\nabla \times G(\bm{s},\bm{r'},\omega))_{m j} \ , \label{eqn:E_correlation_PML_contribution}
\end{align}
which are the additional fluctuations caused by the gain components of the resonator and the PMLs. The correlations based on the term $C_\mathrm{gain}^\mathrm{PML}(\bm{r},\bm{r}')$ are not physical, and we must therefore formulate our modal theory in such a way that it reproduces the correct correlation function without this term.

To achieve this, we redefine the gain and loss operators in Eqs.~\eqref{eqn:QNM_operataors_unsymmetrized_e} and \eqref{eqn:QNM_operataors_unsymmetrized_m} by including the gain components of the PML region $\Omega_\mathrm{PML}$ in the loss operators:
\begin{align}\label{eqn:QNM_operators_unsymmetrized_e_L_2}
    \hat{\xi}_{n,\mathrm{e},\mathrm{L}} = + \mathrm{i} \mu_0 \int_0^\infty\mathrm{d}\omega\, \omega^2 A_n(\omega)\int_\Omega\mathrm{d}^3 r\, \bigg[\theta(\tilde{\varepsilon}_{\mathrm{i},j}(\bm{r},\omega)) \notag\\ 
    \ \ \ \ \sqrt{\frac{\hbar\varepsilon_0}{\pi}|\tilde{\varepsilon}_{\mathrm{i},j}(\bm{r},\omega) |} \bm{e}_j\cdot \tilde{\bm{E}}_n(\bm{r}) \hat{f}_{\mathrm{e},j}(\bm{r},\omega) \bigg]\notag \\
    + \mathrm{i} \mu_0 \int_0^\infty\mathrm{d}\omega\, \omega^2 A_n(\omega)\int_{\Omega_\mathrm{PML}}\mathrm{d}^3 r\, \bigg[ \theta(-\tilde{\varepsilon}_{\mathrm{i},j}(\bm{r},\omega)) \notag\\ 
    \sqrt{\frac{\hbar\varepsilon_0}{\pi}|\tilde{\varepsilon}_{\mathrm{i},j}(\bm{r},\omega) |} \bm{e}_j\cdot \tilde{\bm{E}}_n(\bm{r}) \hat{f}_{\mathrm{e},j}^{\dagger}(\bm{r},\omega) \bigg] \notag \\
    = \hat{\tilde{\xi}}_{n,\mathrm{e},\mathrm{L}} + \hat{\tilde{\xi}}^\dagger_{n,\mathrm{e},\mathrm{PML}} \ , 
\end{align}
\begin{align}\label{eqn:QNM_operators_unsymmetrized_e_G_2}
    \hat{\xi}^\dagger_{n,\mathrm{e},\mathrm{G}} =  \mathrm{i} \mu_0 \int_0^\infty\mathrm{d}\omega\, \omega^2 A_n(\omega)\int_{\Omega\setminus\Omega_\mathrm{PML}}\mathrm{d}^3 r\, \bigg[\theta(-\tilde{\varepsilon}_{\mathrm{i},j}(\bm{r},\omega)) \notag\\ 
    \sqrt{\frac{\hbar\varepsilon_0}{\pi}|\tilde{\varepsilon}_{\mathrm{i},j}(\bm{r},\omega) |} \bm{e}_j\cdot \tilde{\bm{E}}_n(\bm{r}) \hat{f}_{\mathrm{e},j}^{\dagger}(\bm{r},\omega) \bigg] \ , 
\end{align}
\begin{align}\label{eqn:QNM_operators_unsymmetrized_m_L_2}
    \hat{\xi}_{n,\mathrm{m},\mathrm{L}} = + \mathrm{i} \mu_0 \int_0^\infty\mathrm{d}\omega\, \omega A_n(\omega)\int_\Omega\mathrm{d}^3 r\, \bigg[\theta(-\tilde{\kappa}_{\mathrm{i},j}(\bm{r},\omega)) \notag\\ 
    \sqrt{\frac{\hbar}{\pi \mu_0}|\tilde{\kappa}_{\mathrm{i},j}(\bm{r},\omega) |} \bm{e}_j\cdot (\nabla\times\tilde{\bm{E}}_n(\bm{r})) \hat{f}_{\mathrm{m},j}(\bm{r},\omega) \bigg] \notag \\
    + \mathrm{i} \mu_0 \int_0^\infty\mathrm{d}\omega\, \omega A_n(\omega)\int_{\Omega_\mathrm{PML}}\mathrm{d}^3 r\, \bigg[\theta(\tilde{\kappa}_{\mathrm{i},j}(\bm{r},\omega)) \notag\\ 
    \sqrt{\frac{\hbar}{\pi \mu_0}|\tilde{\kappa}_{\mathrm{i},j}(\bm{r},\omega) |} \bm{e}_j\cdot (\nabla\times\tilde{\bm{E}}_n(\bm{r})) \hat{f}_{\mathrm{m},j}^{\dagger}(\bm{r},\omega) \bigg] \notag \\
    = \hat{\tilde{\xi}}_{n,\mathrm{m},\mathrm{L}} + \hat{\tilde{\xi}}^\dagger_{n,\mathrm{m},\mathrm{PML}} \ , 
\end{align}
\begin{align}\label{eqn:QNM_operators_unsymmetrized_m_G_2}
    \hat{\xi}^\dagger_{n,\mathrm{m},\mathrm{G}} = + \mathrm{i} \mu_0 \int_0^\infty\mathrm{d}\omega\, \omega A_n(\omega)\int_{\Omega \setminus\Omega_\mathrm{PML}}\mathrm{d}^3 r\, \bigg[\theta(\tilde{\kappa}_{\mathrm{i},j}(\bm{r},\omega)) \notag\\ 
    \sqrt{\frac{\hbar}{\pi \mu_0}|\tilde{\kappa}_{\mathrm{i},j}(\bm{r},\omega) |} \bm{e}_j\cdot (\nabla\times\tilde{\bm{E}}_n(\bm{r})) \hat{f}_{\mathrm{m},j}^{\dagger}(\bm{r},\omega) \bigg] \ .
\end{align}
These loss operators thus satisfy the commutation relations
\begin{align}\label{eqn:P_e_L}
    \left[ \hat{\xi}_{n,\mathrm{e},\mathrm{L}}, \hat{\xi}^\dagger_{n',\mathrm{e},\mathrm{L}} \right]  = \frac{\hbar \mu_0}{\pi} \int_0^\infty\mathrm{d}\omega\, \omega^2 A_n(\omega) A^*_{n'}(\omega) \notag \\
    \bigg[+ \int_\Omega\mathrm{d}^3 r\, 
     \theta(\tilde{\varepsilon}_{\mathrm{i},j}(\bm{r},\omega)) 
    \frac{\omega^2}{c^2} |\tilde{\varepsilon}_{\mathrm{i},j}(\bm{r},\omega)| E_{n,j}(\bm{r})E_{n', j}^*(\bm{r}) \notag \\
    - \int_{\Omega_\mathrm{PML}} \mathrm{d}^3 r\, 
     \theta(-\tilde{\varepsilon}_{\mathrm{i},j}(\bm{r},\omega)) 
    \frac{\omega^2}{c^2} |\tilde{\varepsilon}_{\mathrm{i},j}(\bm{r},\omega)| E_{n,j}(\bm{r})E_{n', j}^*(\bm{r})\bigg]  \notag \\
     = \tilde{P}^{\mathrm{e},\mathrm{L}}_{n n'} - \tilde{P}^{\mathrm{e},\mathrm{PML}}_{n' n}= P^{\mathrm{e},\mathrm{L}}_{n n'}  \ , 
\end{align}
\begin{align}\label{eqn:P_m_L}
    \left[ \hat{\xi}_{n,\mathrm{m},\mathrm{L}}, \hat{\xi}^\dagger_{n',\mathrm{m},\mathrm{L}} \right] = P^{\mathrm{m},\mathrm{L}}_{n n'} = \frac{\hbar \mu_0}{\pi} \int_0^\infty\mathrm{d}\omega\, \omega^2 A_n(\omega) A^*_{n'}(\omega) \notag \\
    \bigg[+\int_\Omega\mathrm{d}^3 r\, 
     \theta(-\tilde{\kappa}_{\mathrm{i},j}(\bm{r},\omega)) 
     |\tilde{\kappa}_{\mathrm{i},j}(\bm{r},\omega)| \notag\\(\nabla\times\bm{E}_n(\bm{r}))_j (\nabla \times\bm{E}_{n'}^*(\bm{r}) )_j \notag \\
     -\int_{\Omega_\mathrm{PML}}\mathrm{d}^3 r\, 
     \theta(\tilde{\kappa}_{\mathrm{i},j}(\bm{r},\omega)) 
     |\tilde{\kappa}_{\mathrm{i},j}(\bm{r},\omega)| \notag \\(\nabla\times\bm{E}_n(\bm{r}))_j (\nabla \times\bm{E}_{n'}^*(\bm{r}) )_j \bigg] \notag \\ 
     = \tilde{P}^{\mathrm{m},\mathrm{L}}_{n n'} - \tilde{P}^{\mathrm{m},\mathrm{PML}}_{n' n}= P^{\mathrm{e},\mathrm{L}}_{n n'}  \ , 
\end{align}
and the gain operators satisfy
\begin{align}
    \left[ \hat{\xi}_{n,\mathrm{e},\mathrm{G}}, \hat{\xi}^\dagger_{n',\mathrm{e},\mathrm{G}} \right] = P^{\mathrm{e},\mathrm{G}}_{n n'} = \frac{\hbar \mu_0}{\pi} \int_0^\infty\mathrm{d}\omega\, \omega^2 A_n^*(\omega) A_{n'}(\omega) \notag \\
    \int_{\Omega\setminus\Omega_\mathrm{PML}} \mathrm{d}^3 r\, \theta(-\tilde{\varepsilon}_{\mathrm{i},j}(\bm{r},\omega)) 
    \frac{\omega^2}{c^2} |\tilde{\varepsilon}_{\mathrm{i},j}(\bm{r},\omega)| E^*_{n,j}(\bm{r})E_{n', j}(\bm{r})
\end{align}
\begin{align}
    \left[ \hat{\xi}_{n,\mathrm{m},\mathrm{G}}, \hat{\xi}^\dagger_{n',\mathrm{m},\mathrm{G}} \right] = P^{\mathrm{m},\mathrm{G}}_{n n'} =  \frac{\hbar \mu_0}{\pi} \int_0^\infty\mathrm{d}\omega\, \omega^2 A_n^*(\omega) A_{n'}(\omega) \notag \\
    \int_{\Omega\setminus\Omega_\mathrm{PML}}\mathrm{d}^3 r\, 
     \theta(\tilde{\kappa}_{\mathrm{i},j}(\bm{r},\omega)) 
     |\tilde{\kappa}_{\mathrm{i},j}(\bm{r},\omega)| \notag \\
     (\nabla\times\bm{E}^*_n(\bm{r}))_j (\nabla \times\bm{E}_{n'}(\bm{r}) )_j 
\end{align}

All the other commutators vanish due to the bosonic commutation relations of the operators $\hat{\bm{f}}_\lambda(\bm{r},\omega)$ and the distinct loss/gain regions. 
Since the commutation relations of these modal operators are non-bosonic, we have to apply symmetrization transformations separately to the electric/magnetic and loss/gain components in order to ascertain bosonic modal operators~\cite{franke2019quantization,lowdin1950non}:
\begin{align}\label{eqn:symmetrized_QNM_operators}
    \hat{a}_{n, \mathrm{e/m}, \mathrm{L/G}}  = \left(P^{\mathrm{e/m},\mathrm{L/G}} \right)^{-\frac{1}{2}}_{n n'} \hat{\xi}_{n', \mathrm{e/m}, \mathrm{L/G}} \, .
\end{align}
Note that the matrices of the new loss operators $P^{\mathrm{e/m},\mathrm{L}}_{n n'}$ now have negative contributions from the gain components of the PMLs, as can be seen in Eqs.~\eqref{eqn:P_e_L} and \eqref{eqn:P_m_L}. However, as the loss is much more dominant in the PMLs, we will still assume that these matrices are positive-definite, ensuring the existence of the inverses and square roots. These new operators $\hat{a}_{n, \mathrm{e/m}, \mathrm{L/G}}$ satisfy bosonic commutation relations:
\begin{align}
    \left[ \hat{a}_{n, \lambda, X}, \hat{a}^\dagger_{n', \lambda', X'} \right] = \delta_{n n'} \delta_{\lambda \lambda'} \delta_{X X'} \ ,
\end{align}
with $\lambda = \mathrm{e,m}$ and $X = \mathrm{L,G}$. Due to the different commutators, these operators correspond to differently symmetrized modes, which are superpositions of the original modes $\tilde{\bm{E}}_n(\bm{r)}$:
\begin{align}\label{eqn:symmetrized_modes_general}
    \tilde{\bm{E}}^{\mathrm{sym}}_{n,\lambda, \mathrm{L}}(\bm{r}) &= \left(P^{\lambda,\mathrm{L}} \right)^{\frac{1}{2}}_{n' n}\tilde{\bm{E}}_{n'}(\bm{r}) \, , \\
    \tilde{\bm{E}}^{\mathrm{sym}}_{n,\lambda, \mathrm{G}}(\bm{r}) &= \left(P^{\lambda,\mathrm{G}} \right)^{\frac{1}{2}}_{n n'}\tilde{\bm{E}}_{n'}(\bm{r}) \, . \label{eqn:symmetrized_modes_general_gain}
\end{align}
The electric field operator can then be written as
\begin{align}\label{eqn:E_field_operator_sym}
    \hat{\bm{E}}^{(+)}(\bm{r}) = \sum_{n,\lambda} \tilde{\bm{E}}^{\mathrm{sym}}_{n,\lambda, \mathrm{L}}(\bm{r})\hat{a}_{n, \lambda, \mathrm{L}} + \tilde{\bm{E}}^{\mathrm{sym}}_{n,\lambda, \mathrm{G}}(\bm{r})\hat{a}^\dagger_{n, \lambda, \mathrm{G}} \, .
\end{align}
From the definitions of the modal operators in Eqs.~\eqref{eqn:QNM_operators_unsymmetrized_e_L_2} to \eqref{eqn:QNM_operators_unsymmetrized_m_G_2}, it is apparent that the loss operators do not have the same vacuum state $\ket{0}$ as the continuum operators $\hat{\bm{f}}_\lambda(\bm{r},\omega)$, as we have included the gain part of the PMLs in the loss operators ($\hat{a}_{n,\mathrm{e/m},\mathrm{L}} \ket{0} \neq 0$). Nevertheless, we adopt it as the vacuum state of the modal operators ($\hat{a}_{n,\mathrm{e/m},\mathrm{L}} \ket{0} = 0$) in order to eliminate the unphysical contribution of the gain components of the PMLs from the vacuum correlation function in Eq.~\eqref{eqn:E_correlation_PML_contribution}, and use it to construct a modal bosonic Fock space $\mathcal{F}$, in which the $k$-photon Fock states can be generated from $\ket{0}$ in the usual manner:
\begin{align}
    \ket{k_{n,\lambda, X}} = \frac{1}{\sqrt{k_{n,\lambda, X}!}} \left( \hat{a}^\dagger_{n,\lambda, X} \right)^k \ket{0} \ . 
\end{align}
The commutator calculated from Eq.~\eqref{eqn:E_field_operator_sym} is clearly equal to the one calculated from the exact theory as $P_{n n'} = P^{\mathrm{e},\mathrm{L}}_{n n'} + P^{\mathrm{m},\mathrm{L}}_{n n'} - P^{\mathrm{e},\mathrm{G}}_{n' n} - P^{\mathrm{e},\mathrm{G}}_{n' n}$. Furthermore, if we calculate the vacuum correlation function from Eq.~\eqref{eqn:E_field_operator_sym} in the constructed Fock space, we now obtain
\begin{align}
    &\langle \hat{E}_i(\bm{r})\hat{E}_j(\bm{r}') \rangle \\
    &= \sum_{n, n'} (P^{\mathrm{e},\mathrm{L}}_{n n'} + P^{\mathrm{m},\mathrm{L}}_{n n'} + P^{\mathrm{e},\mathrm{L}}_{n' n} + P^{\mathrm{m},\mathrm{L}}_{n' n}   ) \tilde{E}_{n,i}(\bm{r})\tilde{E}^*_{n',j}(\bm{r}') \notag \\
    &= \frac{\hbar \mu_0}{\pi} \int_0^\infty \mathrm{d}\omega\, \omega^2 \bigg[ \notag \\
    &\int_{\Omega\setminus \Omega_\mathrm{PML}} \frac{\omega^2}{c^2} G^\dagger_{i m}(\bm{s},\bm{r},\omega) |\tilde{\varepsilon}_{\mathrm{i},m}(\bm{s},\omega)| G_{m j}(\bm{s},\bm{r}',\omega) \notag \\
  &+ (\nabla \times G(\bm{s},\bm{r},\omega))^\dagger_{i m} |\tilde{\kappa}_{\mathrm{i},m}(\bm{s},\omega)|(\nabla \times G(\bm{s},\bm{r'},\omega))_{m j}  \notag \\
  +&\int_{\Omega_\mathrm{PML}} \frac{\omega^2}{c^2} G^\dagger_{i m}(\bm{s},\bm{r},\omega) \tilde{\varepsilon}_{\mathrm{i},m}(\bm{s},\omega) G_{m j}(\bm{s},\bm{r}',\omega) \notag \\
  &- (\nabla \times G(\bm{s},\bm{r},\omega))^\dagger_{i m} \tilde{\kappa}_{\mathrm{i},m}(\bm{s},\omega)(\nabla \times G(\bm{s},\bm{r'},\omega))_{m j} \bigg] \notag \ .
\end{align}
Note that in contrast to the exact vacuum correlation function in the presence of the PMLs [see Eq.~\eqref{eqn:vacuum_correlation_PMLs}], the moduli of the material parameters do not appear in the integral over the PML region. By using the identity in Eq.~\eqref{eqn:Im_Greens}, we can write the correlation function as
\begin{align}
    &\langle \hat{E}_i(\bm{r})\hat{E}_j(\bm{r}') \rangle = \frac{\hbar \mu_0}{\pi} \int_0^\infty \mathrm{d}\omega\, \omega^2 \bigg[ \Im(G(\bm{r},\bm{r}',\omega))_{i j} \\
    &+2\int_{\Omega\setminus \Omega_\mathrm{PML}}\mathrm{d}^3 r\, \frac{\omega^2}{c^2}\theta(-\tilde{\varepsilon}_{\mathrm{i},m} (\bm{r},\omega) ) \notag \\
    &\hspace{2cm}\cdot G^\dagger_{i m}(\bm{s},\bm{r},\omega) |\tilde{\varepsilon}_{\mathrm{i},m}(\bm{s},\omega)| G_{m j}(\bm{s},\bm{r}',\omega) \notag \\
    &\hspace{2.7cm}+ \theta(\tilde{\kappa}_{\mathrm{i},m} (\bm{r},\omega) ) \notag \\
    &\cdot (\nabla \times G(\bm{s},\bm{r},\omega))^\dagger_{i m} |\tilde{\kappa}_{\mathrm{i},m}(\bm{s},\omega)|(\nabla \times G(\bm{s},\bm{r'},\omega))_{m j} \bigg] \notag \ ,
\end{align}
showing that it is not altered by the presence of the PMLs and reflects the correlation function of the real system without PMLs. In this way, the modal theory reproduces the correct commutators and vacuum correlation functions of the exact theory without the unphysical contribution from the gain components of the PMLs. The magnetic field operator can be expanded using the same bosonic modal operators:
\begin{align}\label{eqn:B_operator_symmetrized}
    \hat{\bm{B}}^{(+)}(\bm{r}) =  \sum_{n,\lambda} \tilde{\bm{B}}^{\mathrm{sym}}_{n,\lambda, \mathrm{L}}(\bm{r})\hat{a}_{n, \lambda, \mathrm{L}} + \tilde{\bm{B}}^{\mathrm{sym}}_{n,\lambda, \mathrm{G}}(\bm{r})\hat{a}^\dagger_{n, \lambda, \mathrm{G}} \ , 
\end{align}
with
\begin{align}
    \tilde{\bm{B}}^{\mathrm{sym}}_{n,\lambda, \mathrm{L}}(\bm{r}) &= (P^{\lambda,\mathrm{L}})^\frac{1}{2}_{n' n} \frac{1}{\mathrm{i}\tilde{\omega}_{n'}} \nabla \times \tilde{\bm{E}}_{n'}(\bm{r}) \ , \\
    \tilde{\bm{B}}^{\mathrm{sym}}_{n,\lambda, \mathrm{G}}(\bm{r}) &= (P^{\lambda,\mathrm{G}})^\frac{1}{2}_{n n'} \frac{1}{\mathrm{i}\tilde{\omega}_{n'}} \nabla \times \tilde{\bm{E}}_{n'}(\bm{r}) \ .
\end{align}
It should be emphasized that so far, we have treated the most general case of linear, magnetodielectric media, for which the permittivities and permeabilities can exhibit gain contributions, and introduced four operators per mode. In the following, we will discuss three important simplifications and special cases in which the complexity of the modal description can be reduced considerably.

\textbf{1. Combination of electric/magnetic operators}

    The electric and magnetic operators can in principle always be combined as $\hat{\xi}_{n, X} = \hat{\xi}_{n,\mathrm{e},X} + \hat{\xi}_{n,\mathrm{m},X}$, even if the electric and magnetic responses exhibit gain in distinct spatial/spectral regions. In this case only two operators per mode are needed, which satisfy the commutation relations
\begin{align}
    \left[ \hat{\xi}_{n, X},\hat{\xi}^\dagger_{n', X} \right] = P^\mathrm{X}_{n n'} = P^{\mathrm{e},X}_{n n'} + P^{\mathrm{m},X}_{n n'} \ . 
\end{align}
After the corresponding symmetrization, similar to Eq.~\eqref{eqn:symmetrized_QNM_operators}, the expanded electric field operator would read 
\begin{align}
    \hat{\bm{E}}^{(+)}(\bm{r}) = \sum_n \tilde{\bm{E}}^{\mathrm{sym}}_{n, \mathrm{L}}(\bm{r}) \hat{a}_{n,\mathrm{L}} + \tilde{\bm{E}}^{\mathrm{sym}}_{n, \mathrm{G}}(\bm{r}) \hat{a}^\dagger_{n,\mathrm{G}} \ , 
\end{align}
in which the symmetrized modes are defined as in Eqs.~\eqref{eqn:symmetrized_modes_general} and \eqref{eqn:symmetrized_modes_general_gain}, with the replacements $P^{\lambda, X} \rightarrow P^{X}$. Even though this simplification can always be made, we introduced distinct electric/magnetic modal operators above as they may provide additional physical information in the case of magnetodielectric systems, where for instance, the modes might be hybrid in character with plasmonic and magnonic components. When studying the decay of QEs in such systems, the splitting of the modal operators would also provide information on whether the QEs dominantly excite plasmonic or magnonic polaritons, as the operators $\hat{\xi}_{n,\mathrm{e/m},X}$ are superpositions of electric/magnetic, polaritonic, matter-field excitations $\hat{\bm{f}}_\mathrm{e/m}(\bm{r},\omega)$.  

\textbf{2. Combination of gain/loss operators}

If we combine the gain and loss operators as $\hat{\xi}_{n,\lambda} = \hat{\xi}_{n,\lambda,\mathrm{L}} + \hat{\xi}_{n,\lambda,\mathrm{G}}^\dagger$, they would satisfy the commutation relations
\begin{align}
[\hat{\xi}_{n,\lambda}, \hat{\xi}_{n',\lambda}^\dagger ] = P^\lambda_{n n'} = P^{\lambda,\mathrm{L}}_{n n'} -P^{\lambda,\mathrm{G}}_{n' n} \ ,
\end{align}
just as we have seen when we included the gain components of the PMLs in the loss operators of Eqs.~\eqref{eqn:P_e_L} and \eqref{eqn:P_m_L}. However, caution is required at this point, because the vacuum fluctuations and the corresponding pumping of the modes due to the gain components of the resonator have a physical influence that generally cannot be neglected, and which becomes clear in the effective master equation derived in \S~\ref{Sec:Master_Eq}. As discussed in Ref.~\cite{franke2022coupled}, two conditions must be met for the gain and loss operators to be combined. Firstly, the combined matrices $P^\mathrm{\lambda}_{n n'}$ must be positive-definite, which is required for the symmetrization of the modal operators. Secondly, as already pointed out when treating the gain components of the PMLs, the combined operators $\hat{\xi}_{n,\lambda}$ do not have the same vacuum state $\ket{0}$ as the continuum operators $\hat{\bm{f}}_\lambda(\bm{r},\omega)$. Applying the modal number operators (without the contributions from the PML gain components) to the vacuum state leads to $\sum_n\hat{a}^\dagger_{n,\lambda}\hat{a}_{n,\lambda}\ket{0} = \sum_n n^0_{n,\lambda} \ket{0}$ with $\sum_n n^0_{n,\lambda} = \mathrm{Tr}\{P^{\lambda,\mathrm{G} *} (P^{\lambda})^{-1} \}$. Consequently, the vacuum state of the complete system can only be used approximately as a vacuum state of the modal Fock space if $\sum_n n^0_{n,\lambda} \ll 1$, which holds whenever the losses in the system dominate the gain ({\textit{i.e.}, $P^{\lambda,\mathrm{L}}_{n n} \gg P^{\lambda,\mathrm{G}}_{n n}$) and the off-diagonal elements $P^{\lambda,\mathrm{G}}_{n n'}$ are small. In this case, the expanded electric field operator would read
\begin{align}
    \hat{\bm{E}}^{(+)}(\bm{r}) = \sum_n \tilde{\bm{E}}^\mathrm{sym}_{n,\mathrm{e}}(\bm{r}) \hat{a}_{n,\mathrm{e}} + \tilde{\bm{E}}^\mathrm{sym}_{n,\mathrm{m}}(\bm{r}) \hat{a}_{n,\mathrm{m}} \ ,
\end{align}
in which the symmetrized modes are as defined in Eq.~\eqref{eqn:symmetrized_modes_general} with the replacement $P^{\lambda,\mathrm{L}} \rightarrow P^\lambda$. From this electric field operator, we would obtain the vacuum correlation function $\langle \hat{E}_i(\bm{r})\hat{E}_j(\bm{r}') \rangle = ({\hbar \mu_0}/{\pi})\int_0^\infty\mathrm{d}\omega\, \omega^2 \, \Im(G_{i j}(\bm{r},\bm{r}',\omega) ) $, and therefore the quantum effects caused by the presence of the gain regions and embodied by the additional term in Eq.~\eqref{eqn:E_correlation_PML_contribution} are neglected. However, the gain still affects the correlation function as it alters the modes and the Green's function of the system. Such an approach would not be appropriate to describe systems exhibiting strong gain, as in the context of lasing or exceptional points, where the combined $P$-matrix $P^\lambda_{n n'}$ may fail to be positive-definite; in this circumstance, the required mode symmetrization cannot be carried out. It should be understood that such behaviour is not a failure exclusive to the modal formalism, but rather a general failure that reflects the underlying approximations of the approach. Indeed, the case in which the $P$-matrix becomes non-positive-definite is one in which the photonic local density of states, proportional to $\Im(G_{i i}(\bm{r},\bm{r},\omega))$, becomes non-positive for some $\bm{r}$ and $\omega$, which marks a fundamental breakdown of the aforementioned approximation of the vacuum correlation function via the imaginary part of the Green's function, since $\langle \hat{E}_i(\bm{r})\hat{E}_i(\bm{r}) \rangle$
%= ({\hbar \mu_0}/{\pi})\int_0^\infty\mathrm{d}\omega\, \omega^2 \, \Im(G_{i j}(\bm{r},\bm{r}',\omega) )$
can become negative and this is unphysical.
%Under these conditions, $\langle \hat{E}_i(\bm{r})\hat{E}_i(\bm{r}) \rangle = ({\hbar \mu_0}/{\pi})\int_0^\infty\mathrm{d}\omega\, \omega^2 \, \mathrm{Im}(G_{i j}(\bm{r},\bm{r}',\omega) ) $ clearly breaks down and therefore this approximation could not be used anyways.  } 

\textbf{3. Combination of all four operators}

If the electric/magnetic modal operators are combined and we additionally combine the gain/loss operators, only one operator per mode $\hat{\xi}_n = \hat{\xi}_{n,\mathrm{e},\mathrm{L}} + \hat{\xi}_{n,\mathrm{m},\mathrm{L}} + \hat{\xi}^\dagger_{n,\mathrm{e},\mathrm{G}} + \hat{\xi}^\dagger_{n,\mathrm{m},\mathrm{G}}$ is needed, satisfying the commutation relations
\begin{align}
    \left[ \hat{\xi}_n , \hat{\xi}^\dagger_{n'} \right] = P_{n n' } = P^{\mathrm{e},\mathrm{L}}_{n n'} + P^{\mathrm{m},\mathrm{L}}_{n n'} - P^{\mathrm{e},\mathrm{G}}_{n' n} - P^{\mathrm{m},\mathrm{G}}_{n' n} \ ,
\end{align}
and leading to the expansion
\begin{align}
    \hat{\bm{E}}^{(+)}(\bm{r}) = \sum_n \tilde{\bm{E}}^{\mathrm{sym}}_{n}(\bm{r}) \hat{a}_n \ , 
\end{align}
in which the symmetrized modes are defined as in Eq.~\eqref{eqn:symmetrized_modes_general} with the replacement $P^{\lambda, \mathrm{L}} \rightarrow P$. Note that the combined gain and loss matrices $P^\mathrm{X}_{n n'} = P^{\mathrm{e},\mathrm{X}}_{n n'}  + P^{\mathrm{m},\mathrm{X}}_{n n'}$ have to satisfy the same conditions as discussed in the last paragraph. 

It is also worth mentioning that in our approach, the limit of lossless/gainless materials (\textit{i.e.}, $\zeta_\mathrm{i} \rightarrow 0$) can be taken from the beginning, since we map the problem onto a compact domain $\Omega \subset\mathbb{R}^3$. This means that in this case, we can interchange the limit and the integration in the symmetrization matrix elements $P^{\lambda,\mathrm{L}}_{n n'}$, and consequently only the PML region $\Omega_\mathrm{PML}$ contributes, which corresponds to the radiation loss in Ref.~\cite{franke2020fluctuation}.

\section{\label{Sec:Master_Eq}Effective Lindblad Master Equation}

In this section, we derive an effective Lindblad master equation for the electromagnetic modes and the additional coupling to a QE, described as a two-level system (TLS). We will start by considering the dynamics of the modes only in the most general scenario, in which they are characterized by the operators $\hat{a}_{n,\lambda,X}$. In the Heisenberg picture, the dynamics of each modal operator is determined by the Heisenberg equation of motion $\mathrm{d}\hat{O}/\mathrm{d}t =  \mathrm{i} [\hat{H},\hat{O}]/\hbar$, leading to~\cite{franke2022coupled}
\begin{align}\label{Heisenberg_Langevin_1}
    \frac{\mathrm{d} \hat{a}_{n,\lambda,\mathrm{L}}}{\mathrm{d} t} &= \frac{\mathrm{i}}{\hbar} \left[ \hat{H}_\mathrm{eff},\hat{a}_{n,\lambda,\mathrm{L}}  \right] - \chi^{\lambda,\mathrm{L},-}_{n n'}   \hat{a}_{n',\lambda,\mathrm{L}} + \hat{F}_{n,\lambda, \mathrm{L}}(t)   \ , \\
    \frac{\mathrm{d} \hat{a}_{n,\lambda,\mathrm{G}}}{\mathrm{d} t} &= \frac{\mathrm{i}}{\hbar} \left[ \hat{H}_\mathrm{eff},\hat{a}_{n,\lambda,\mathrm{G}}  \right] - \chi^{\lambda,\mathrm{G},-}_{n n'} \hat{a}_{n,\lambda,\mathrm{G}} + \hat{F}_{n,\lambda,\mathrm{G}}(t) \ , \label{Heisenberg_Langevin_2}
\end{align}
with an effective Hamiltonian
\begin{align}\label{eqn:H_eff}
    \hat{H}_\mathrm{eff} = \hbar \chi^{\lambda,\mathrm{L},+}_{n n'} \hat{a}_{n,\lambda,\mathrm{L}}^\dagger \hat{a}_{n',\lambda,\mathrm{L}} -\hbar \chi^{\lambda,\mathrm{G},+}_{n n'} \hat{a}_{n,\lambda,\mathrm{G}}^\dagger \hat{a}_{n',\lambda,\mathrm{G}}  \ , 
\end{align}
and the coupling matrices
\begin{align}\label{chi_mat_1}
    \chi^{\lambda,X,+}_{n n'} &= \frac{1}{2} (P^{\lambda,X})^{-\frac{1}{2}}_{n n''} (\tilde{\omega}_{n''} + \tilde{\omega}_{n'''}^* ) P^{\lambda,X}_{n'' n'''}(P^{\lambda,X})^{-\frac{1}{2}}_{n''' n'} \ , \\
    \chi^{\lambda,X,-}_{n n'} &= \frac{\mathrm{i}}{2} (P^{\lambda,X})^{-\frac{1}{2}}_{n n''} (\tilde{\omega}_{n''} - \tilde{\omega}_{n'''}^* ) P^{\lambda,X}_{n'' n'''}(P^{\lambda,X})^{-\frac{1}{2}}_{n''' n'}  \label{chi_mat_4} \ , 
\end{align}
The $\hat{F}_{n,\lambda,X}(t)$ are additional operator terms that still depend on the continuum operators $\hat{\bm{f}}_\lambda(\bm{r},\omega)$, and which cannot be expressed simply in terms of the modal operators.
From the classical QNM theory, it is clear that the amplitudes of the freely evolving modes are oscillating with the complex frequencies $\tilde{\omega}_n$, as can be seen in Eq.~\eqref{eqn:QNM_coeffs_time_domain}, meaning the expectation values of the modal operators should satisfy $\langle \hat{\xi}_{n,\lambda,\mathrm{L}}(t) \rangle = \langle \hat{\xi}_{\mu,\lambda,\mathrm{L}}(0) \rangle \mathrm{e}^{-\mathrm{i} \tilde{\omega}_n t}$ and $\langle \hat{\xi}_{n,\lambda,\mathrm{G}}(t) \rangle = \langle \hat{\xi}_{n,\lambda,\mathrm{G}}(0) \rangle \mathrm{e}^{+\mathrm{i} \tilde{\omega}^*_n t}$. From this requirement, it is clear that $\langle \hat{F}^{(\dagger)}_{n,\lambda,X}(t)\rangle = 0$. Furthermore, the peaked structure was factored out in the spectral representations of the operators $\hat{F}_{n,\lambda,X}(t)$ due to terms of the form $(\tilde{\omega}_n - \omega ) A_n(\omega) = \varepsilon_0 c^2 /\tilde{\omega}_n $.
We thus make a Markov approximation and treat them as white quantum noise forces induced by the dissipation/gain and that ensure the preservation of the commutation relations; that is, they satisfy $\left[ \hat{F}_{n,\lambda, \mathrm{X}}(t) , \hat{F}^\dagger_{n',\lambda, X}(t')  \right] = 2\chi^{\lambda,X,-}_{n n'} \delta(t - t') $. This procedure is similar to the reaction coordinate mapping, in which the coupling of a TLS to a structured bath having a strongly peaked spectral density is mapped to an interaction with a discrete mode and a residual continuum that can then be treated within the Markov approximation~\cite{binder2018thermodynamics,anto2021capturing}. Note that in the equation for the modal loss operators $\hat{a}_{n,\lambda,\mathrm{L}}$, only a single loss-induced noise force $\hat{F}_{m,\lambda,\mathrm{L}}$ appears that counteracts the loss term described by $\chi_{n n'}^{\lambda,\mathrm{L},-}$, caused by the overall loss in the system and the gain components of the PMLs, which we have included in the loss operators [see Eq.~\eqref{eqn:QNM_operators_unsymmetrized_e_L_2}]. The gain components of the PMLs are thus only treated as a reduction of the loss in the system, as they cannot induce any real physical gain to the modes in terms of pumping. This is different to how the real gain components of the resonator have to be treated if the gain and loss operators are combined, as in this case the gain introduces pumping of the modes~\cite{franke2022coupled}. Where the loss and gain operators are treated separately, the pumping of the modes due to the gain components of the resonator is described by negative-energy contributions in the effective Hamiltonian of Eq.~\eqref{eqn:H_eff}, associated to the modal gain operators.

By invoking a quantum analogue of It{\^{o}}-Stratonovich calculus, following the same procedure as in Refs.\cite{gardiner1985input, franke2019quantization, franke2022coupled}, an effective Markovian master equation, corresponding to the Heisenberg-Langevin equations in Eqs.~\eqref{Heisenberg_Langevin_1} and \eqref{Heisenberg_Langevin_2}, can be obtained in the form
\begin{align}\label{eqn:Master_equation_general}
    \frac{\mathrm{d}\rho}{\mathrm{d}t} = -\frac{\mathrm{i}}{\hbar} \left[\hat{H}_\mathrm{eff}, \rho \right] + \chi^{\lambda,\mathrm{L}, -}_{n n'} \mathcal{D}^{\lambda,\mathrm{L}}_{n n'} \rho + \chi^{\lambda,\mathrm{G}, -}_{n n'} \mathcal{D}^{\lambda,\mathrm{G}}_{n n'} \rho  \ , 
\end{align}
with the same effective Hamiltonian as in Eq.~\eqref{eqn:H_eff} and the Lindblad dissipators
\begin{align}\label{eqn:dissipator_general}
    \mathcal{D}^{\lambda,X}_{n n'} \rho = 2 \hat{a}_{n',\lambda, X}\rho \hat{a}_{n,\lambda, X}^\dagger - \{\rho, \hat{a}_{n,\lambda, X}^\dagger \hat{a}_{n',\lambda, X}\} \ . 
\end{align}

This is the effective master equation that governs the quantum dynamics of the modes for the most general case, featuring decomposed electric/magnetic modal operators. Note that the loss/gain present in the system necessitates a symmetrization of the modes and consequently induces a coupling between the bosonic modes in both the effective Hamiltonian and the Lindblad dissipators (\textit{i.e.}, coherent and dissipative coupling~\cite{wang2020dissipativemag}). The cross terms in the Lindblad dissipators capture interference effects between different decay channels, as the coupled modes dissipate into the same bath. Such cross terms were also derived for a Jaynes-Cummings model where both the QE and the cavity mode dissipate into the same bath, leading to coupled dissipators between the QE and mode operators to describe Fano-like effects~\cite{yamaguchi2021theory}. 

To describe the coupling of the modes to a fixed QE at $\bm{r}_\mathrm{QE} \in \Omega\setminus\Omega_\mathrm{PML}$, modeled as a TLS with eigenstates $\ket{\mathrm{g}}$, $\ket{\mathrm{e}}$ and electric or magnetic transition dipole moment $\bm{d}/\bm{m} = \langle \mathrm{g} | \hat{\bm{d}}/ \hat{\bm{m}} | \mathrm{e} \rangle$, the mQED Hamiltonian $\hat{H}_\mathrm{EM}$ in Eq.~\eqref{eqn:mQED_Hamiltonian_gain} would have to be supplemented by additional terms,
\begin{align}\label{eqn:Ham_couplings_general}
        \hat{H} &= \hat{H}_\mathrm{EM} + \hat{H}_\mathrm{QE} + \hat{H}_\mathrm{int}\\
    &=\hat{H}_\mathrm{EM} + \hbar \omega_\mathrm{QE} \sigma^+ \sigma^- - \hat{\bm{d}}\cdot \hat{\bm{E}}(\bm{r}_\mathrm{QE}) - \hat{\bm{m}}\cdot \hat{\bm{B}}(\bm{r}_\mathrm{QE}) \notag \ , 
\end{align}
in which $\sigma^\pm = \ket{\mathrm{e/g}}\bra{\mathrm{g/e}}$ are the usual TLS raising and lowering operators.
Note that whilst we treat both possibilities in Eq.~\eqref{eqn:Ham_couplings_general} for completeness, the distinct selection rules pertaining to electric and magnetic dipole transitions will ensure that only one of $\bm{d}$ or $\bm{m}$ is non-zero in a given physical application.
By using the modal expansion of the electric field operator in Eq.~\eqref{eqn:E_field_operator_sym} and following the same steps as before, the master equation thereby obtained possesses the same form as in Eq.~\eqref{eqn:Master_equation_general}, but with an effective Hamiltonian
\begin{align}
    &\hat{H}_\mathrm{eff} = \hbar \omega_\mathrm{QE} \sigma^+ \sigma^- \\
    &+\hbar\chi_{n n'}^{\lambda,\mathrm{L},+}\hat{a}_{n,\lambda,\mathrm{L}}^\dagger \hat{a}_{n',\lambda,\mathrm{L}} - \hbar\chi_{n n'}^{\lambda,\mathrm{G},+}\hat{a}_{n,\lambda,\mathrm{G}}^\dagger \hat{a}_{n',\lambda,\mathrm{G}} \notag  \\
    &+ \hbar (g_{n,\lambda, \mathrm{L}} \sigma^+ \hat{a}_{n,\lambda, \mathrm{L}} + \mathrm{h.c.}) + \hbar (g_{n,\lambda, \mathrm{G}} \sigma^+ \hat{a}^\dagger_{n,\lambda, \mathrm{G}} + \mathrm{h.c.}) \ , \notag
\end{align}
where
\begin{align}\label{eqn:QE_couplings_general}
    g_{n,\lambda, X} = -\frac{1}{\hbar} ( \bm{d} \cdot\tilde{\bm{E}}^\mathrm{sym}_{n,\lambda,X}(\bm{r}_\mathrm{QE}) + \bm{m} \cdot\tilde{\bm{B}}^\mathrm{sym}_{n,\lambda,X}(\bm{r}_\mathrm{QE})) \ ,
\end{align}
and we have applied the rotating-wave approximation. It can be appreciated that the gain modal creation operators $\hat{a}^\dagger_{n,\lambda,\mathrm{G}}$ couple to $\sigma^+$ in the rotating-wave approximation since they appear in the positive-frequency parts of the fields in Eqs.~\eqref{eqn:E_field_operator_sym} and \eqref{eqn:B_operator_symmetrized}. If the QE is subject to additional non-radiative decay channels or dephasing processes, the corresponding dissipators should be added to the master equation in Eq.~\eqref{eqn:Master_equation_general}. Fig.~\ref{Fig_Two}(a) offers a schematic representation of the quantum description for an exemplary resonator-QE system, based on the master equation in this general setting, where the action of the operators and all the different couplings and decay channels are labeled. Note that even though the master equation does not contain coupled dissipators between the QE and mode operators, as derived in Ref.~\cite{yamaguchi2021theory}, it is in principle capable of describing coherent Fano interference effects occurring in the QE decay, as the coupling of the QE with the complete set of modes captures its interaction with the electromagnetic environment completely. Such cross terms would appear if a subset of the modes were traced out, as both the QE and the remaining modes couple to these modes.

If the electric and magnetic modal operators are combined, the derivation of the master equation would proceed in a similar manner as before, with the result having the same form as Eq.~\eqref{eqn:Master_equation_general}, but with the replacements $\chi^{\lambda,X,\pm} \rightarrow \chi^{X,\pm}$, which are exactly defined as in Eqs.~\eqref{chi_mat_1} and \eqref{chi_mat_4}, but with the replacement $P^{\lambda,X} \rightarrow P^X$. 

If additionally the gain and loss operators can be combined, the Heisenberg equation for the combined modal operators $\hat{a}_n$ can be written as
\begin{align}
    \frac{\mathrm{d} \hat{a}_n}{\mathrm{d} t} = \frac{\mathrm{i}}{\hbar} \left[ \hat{H}_\mathrm{eff},\hat{a}_n \right] - \tilde{\chi}_{n n'}^{\mathrm{L},-} \hat{a}_{n'} + \hat{F}_{n,\mathrm{L}}(t) + \hat{F}^\dagger_{n,\mathrm{G}}(t) \ , 
\end{align}
with 
\begin{align}
    \hat{H}_\mathrm{eff} = \hbar \chi_{n n'}^{+} \hat{a}^\dagger_n \hat{a}_{n'} \ ,
\end{align}
and the matrices
\begin{align}
    \chi_{n n'}^{+} &= \frac{1}{2} P^{-\frac{1}{2}}_{n n''} ( \tilde{\omega}_{n''} + \tilde{\omega}_{n'''}^* ) P_{n'' n'''} P^{-\frac{1}{2}}_{n''' n'} \ , \\
     \tilde{\chi}^{X,-}_{n n'} &= \frac{\mathrm{i}}{2} P^{-\frac{1}{2}}_{n n''} (\tilde{\omega}_{n''} - \tilde{\omega}_{n'''}^*)P^{X}_{n'' n'''} P^{-\frac{1}{2}}_{n''' n'} \ , 
\end{align}
The corresponding master equation thus has the form
\begin{align}
     \frac{\mathrm{d}\rho}{\mathrm{d}t} = -\frac{\mathrm{i}}{\hbar} \left[ H_\mathrm{eff}, \rho \right] + \tilde{\chi}_{n n'}^{\mathrm{L},-} \mathcal{D}^\mathrm{L}_{n n'}\rho +  \tilde{\chi}_{n n'}^{\mathrm{G},-} \tilde{\mathcal{D}}^{\mathrm{G}}_{n n'}\rho \ , \label{eqn:Master_eq_combined_operators}
\end{align}
in which the dissipators $\mathcal{D}^\mathrm{L}_{n n'}$ are defined just as in Eqs.~\eqref{eqn:dissipator_general}, and $\tilde{\mathcal{D}}^{\mathrm{G}}_{n n'}$ as 
\begin{align}
    \tilde{\mathcal{D}}^{\mathrm{G}}_{n n'} = 2 \hat{a}_n^\dagger \rho \hat{a}_{n'} - \{ \rho, \hat{a}_{n'} \hat{a}_n^\dagger \} \ .
\end{align}
In this case, any gain of the resonator is completely described by incoherent pumping terms (proportional to $\tilde{\chi}^{\mathrm{G},+}_{n n'}$), rather than negative-energy contributions in the effective Hamiltonian. The coupling constants for the interaction with a QE would have the same form as in Eq.~\eqref{eqn:QE_couplings_general}, but with the accordingly symmetrized modes $\tilde{\bm{E}}^\mathrm{sym}_n(\bm{r}_\mathrm{QE}) / \tilde{\bm{B}}^\mathrm{sym}_n(\bm{r}_\mathrm{QE}) $. A schematic representation of the quantum description in this simplified setting is shown in Fig.~\ref{Fig_Two}(b). 
In the case of a resonator with a Drude-Lorentz permittivity and non-dispersive permeability, we can also immediately use the formula obtained from the classical QNM theory in Eq.~\eqref{eqn:QNM_coeffs_time_domain} to describe the driving of the modes by the classical coherent field $\bm{E}_\mathrm{b}(\bm{r},t)$. In the quantum description, we would introduce an additional driving term $\hat{H}_\mathrm{drive}(t)$ in the effective Hamiltonian given by
\begin{align}
    \hat{H}_\mathrm{drive}(t) = \mathrm{i}\hbar P^{-\frac{1}{2}}_{n n'} D_{n'}(t) \hat{a}_n^\dagger + \mathrm{h. c.} \ . 
\end{align}
In practice, an important issue surrounding the numerical solution of the quantum master equation, Eq.~\eqref{eqn:Master_equation_general} or Eq.~\eqref{eqn:Master_eq_combined_operators}, is the exponential scaling of the Hilbert space dimension with the number of modes retained to ensure effective completeness, which may quickly render quantum dynamic simulations infeasible unless suitably optimized, numerical linear algebra and parallel computation strategies are leveraged.

\begin{figure}
    \centering
    \includegraphics[width=1\linewidth]{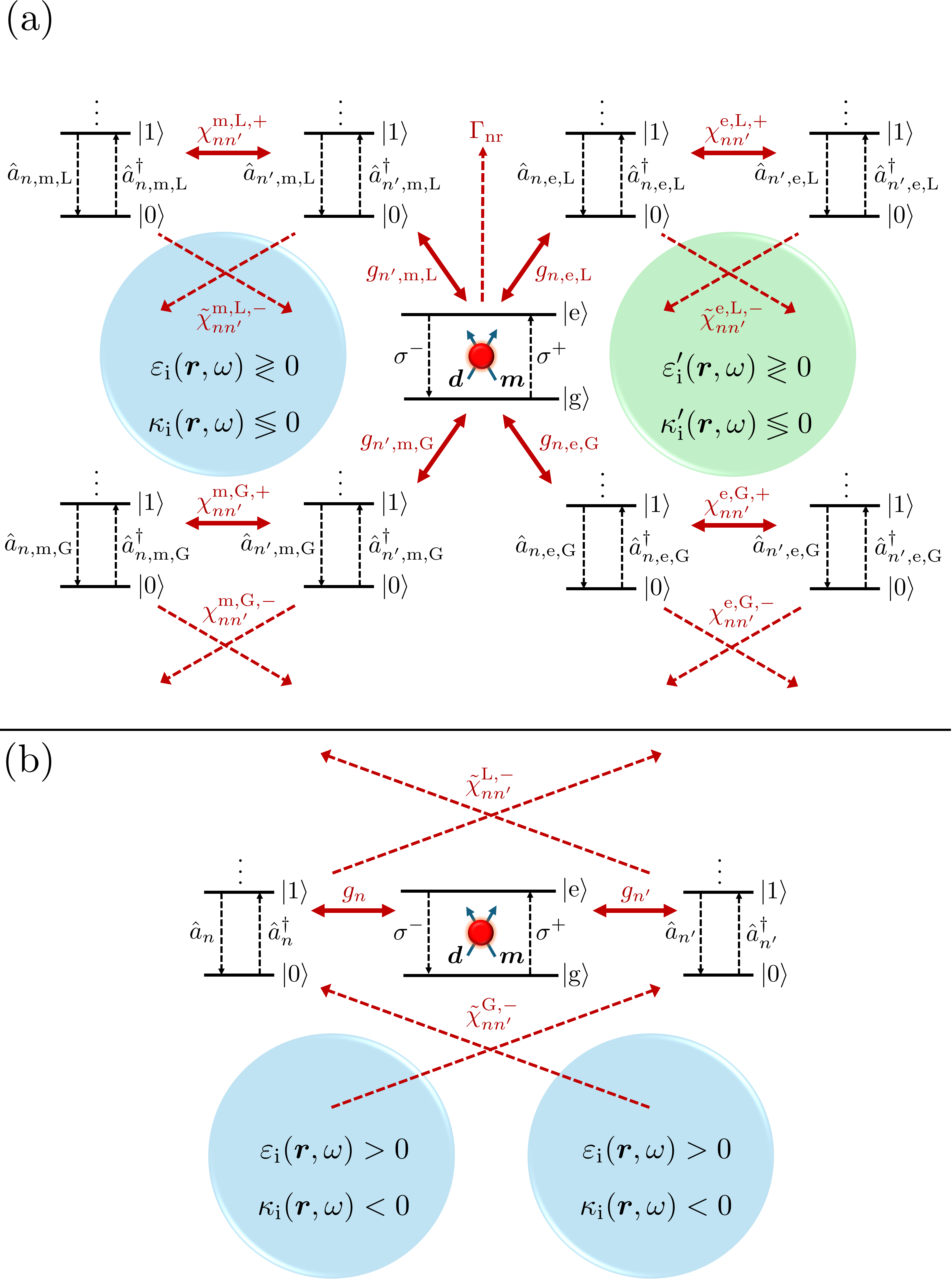}
    \caption{Schematic representation of the quantum description of an electric/magnetic dipolar QE, modeled as a TLS, interacting with an electromagnetic resonator using PMLs. (a) Most general scenario, in which the resonator could have electric and magnetic gain components in distinct spatial or spectral regions (\textit{e.g.}, a pair of coupled, spherical resonators composed of different materials). In this case, every eigenmode of the resonator (both QNMs and PML modes) is described by four bosonic operators $\hat{a}_{n,\lambda,X}$ with $\lambda= \mathrm{e, m}$ and $X = \mathrm{L, G}$, corresponding to differently symmetrized modes according to Eqs.~\eqref{eqn:symmetrized_modes_general} and \eqref{eqn:symmetrized_modes_general_gain}. The presence of gain/loss in the system necessitates a symmetrization and the consequent coupling between the symmetrized modes described by $\chi_{n n'}^{\lambda,X,+}$ in the effective Hamiltonian of Eq.~\eqref{eqn:H_eff}. The losses of the system are described by Lindblad dissipators proportional to $\chi_{n n'}^{\lambda,\mathrm{L},-}$, as given in Eqs.~\eqref{eqn:dissipator_general}, which also involve coupling between the symmetrized modes. The gain components of the resonator are described by negative-energy contributions in the effective Hamiltonian. All the different bosonic modes couple to the QE with the coupling constants $g_{n,\lambda,X}$, as given in Eq.~\eqref{eqn:QE_couplings_general}. Additional non-radiative decay or dephasing processes of the QE, here labeled by $\Gamma_\mathrm{nr}$, can be treated in the usual way by adding corresponding Lindblad terms to the master equation. (b) Scenario in which the electric/magnetic and gain/loss modal operators can be combined intos a single operator per mode $\hat{a}_n$ (\textit{e.g.}, two spheres composed of passive, magnetodielectric media). In this case, the gain of the resonator is described by incoherent pumping terms proportional to $\tilde{\chi}_{n n'}^{\mathrm{G},-}$.   }
    \label{Fig_Two}
\end{figure}

\section{\label{Sec:Numerical_Examples} Numerical Examples}

Having presented the general formalism in the preceding section, we now demonstrate the predictive capabilities of our approach by investigating the spontaneous emission of a QE in two structured, dielectric environments. In \S~\ref{Sec:1D_example}, we consider a 1D, half-open cavity due to its simple mode structure and analytical tractability. As the PMLs in 1D do not exhibit any gain components, we also briefly study a 3D spherical cavity in \S~\ref{Sec:3D_example}, where the proper elimination of the corresponding PML gain contributions enables accurate predictions to be ascertained.

\subsection{\label{Sec:1D_example} 1D Half-open Cavity}

%Having presented the general formalism in the preceding sections, we now demonstrate the predictive capabilities of our approach in a simple, 1D example. 

As schematized in Fig.~\ref{fig:1D_Cavity}(a), we consider a half-open cavity comprising a homogeneous, diamond sample of relative permittivity $\varepsilon = 5.7$ and length $L = 200 \, \textrm{nm}$, with a perfect electrical conductor boundary placed at $x=0$ (corresponding to a Dirichlet boundary condition imposed at this point). In spite of its seeming simplicity, our choice of resonator provides a fairly stringent test of the theory, exhibiting QNMs with very low quality factors and spectrally overlapping resonance features in their coupling with QEs, a setting where phenomenological models based on normal modes can become inaccurate. Of course, our approach is expected to perform just as well with the more widely used, high-finesse Fabry-P{\'e}rot cavity designs of traditional quantum optics.

We assume that the diamond hosts a single QE at some location $x_\mathrm{QE}$ along the $x$ axis, featuring a pair of levels that are coupled by the electric dipole interaction with the cavity field.
%with transition frequency $\omega_\mathrm{QE}$
%with a vertically oriented transition dipole moment ${\bm d}$,
%at some location $x_\mathrm{QE}$ along the $x$ axis.
%characterized by its transition frequency $\omega_\mathrm{QE}$ and a vertically oriented transition dipole moment ${\bm d}$.
Such a QE could be realized in the form of a color center in the diamond lattice, and a variety of such optically active defects have been explored for quantum applications~\cite{aharonovich2014diamond}. For the purpose of our numerical tests, we shall here consider a QE having the transition energy of a nitrogen-vacancy center, namely $\omega_\mathrm{QE} = 1.95\,\mathrm{eV}$ ($637\,\mathrm{nm}$), and a vertically oriented transition dipole moment ${\bm d}$ with controllable magnitude $|{\bm d}|$.
%allowing us to make predictions and explore their accuracy for both the weak and strong coupling regimes.
%The setup is sketched in Fig.~\ref{fig:1D_Cavity_1}(a).
%for such a system, the Green's function can be determined analytically.

A 1D formulation of mQED can be derived for systems that are invariant in two spatial dimensions (say, the $y$ and $z$ directions), and can be used to study the interaction of QEs with linearly polarized radiation propagating in the $x$ direction~\cite{gruner1996mqed}. In this case, we make the replacements $\hat{\bm{E}}(\bm{r}) \rightarrow \hat{E}(x) = \hat{E}_y(x)$, ${\bm{d}} \rightarrow d = d_y$ and $\hat{\bm{f}}(\bm{r},\omega) \rightarrow \hat{f}(x,\omega) =  \hat{f}_y(x,\omega)$. The $y$ component of the noise current density is then given by
\begin{align}
    \hat{j}_{\mathrm{F}}(x,\omega) =  \hat{j}_{\mathrm{F},y}(x,\omega) &= \omega \sqrt{\frac{\hbar \varepsilon_0}{\pi A} \tilde{\varepsilon}(x,\omega) } \hat{f}_\mathrm{e}(x,\omega) \\
    &+ \partial_x \left( \sqrt{- \frac{\hbar}{\pi \mu_0 A} \tilde{\kappa}(x,\omega) } \hat{f}_\mathrm{m}(x,\omega)    \right) \ , \notag 
\end{align}
in which $A$ is the normalization area in the $yz$ plane, and we adopt 1D material parameters $\tilde{\zeta}(x,\omega) = \tilde{\zeta}_{y y}(x,\omega)$ appropriate to a single PML in the $x$ direction, given by
\begin{align}
    \tilde{\zeta}(x,\omega) = \zeta(x,\omega) \left( 1 + \frac{\mathrm{i}}{\omega + \mathrm{i}\eta} \sigma_x(x) \right) \ , 
\end{align}
as can be obtained from Eqs.~\eqref{Eq:PML_Jacobian} and \eqref{eqn:PML_parameter_transformation} with $\sigma_{y,z} = 0$. In the 1D case, the PMLs are purely lossy, as waves with normal incidence experience only their loss components. In our example, we choose PMLs with a step-like absorption profile, $\sigma_x(x) = 2.5\,\mathrm{eV} \cdot\theta(x-L)$ (\textit{i.e.}, they are directly adjacent to the resonator), a thickness of $20 L$ and $\eta = 0.1\,\mathrm{eV}$. It is worth noting that this 1D problem admits an analytical solution for the QNMs and Green's function directly from the boundary and continuity conditions, thus obviating the need for numerical solution of the 1D Helmholtz equation. As such, spurious numerical reflections that would otherwise arise from a spatial discretization do not occur, and a simple, step-like absorption profile, rather than a gradually increasing one, suffices.
%as numerical reflections due to spatial discretization do not occur.  

The formulae for the $P$-matrices all have the same form as in the 3D case, but with the replacements mentioned above and $\nabla \times \rightarrow \partial_x$:
\begin{align}
    &P^\mathrm{L}_{n n'} = \frac{\hbar \mu_0}{\pi A} \int_0^\mathrm{\infty} \mathrm{d}\omega\, \omega^2 A_n(\omega) A_{n'}^*(\omega)  \label{eqn:P_integral_1D} \\ 
    &\int\mathrm{d}s\, \frac{\omega^2}{c^2} \tilde{\varepsilon}_\mathrm{i}(s,\omega) \tilde{E}_n(s) \tilde{E}^*_{n'}(s) - \tilde{\kappa}_\mathrm{i}(s,\omega) \partial_s \tilde{E}_n(s) \partial_s \tilde{E}^*_{n'}(s) \notag \ , 
\end{align}
where we have combined the electric and magnetic modal operators. The corresponding master equation then has the same form as in Eq.~\eqref{eqn:Master_equation_general} with $\chi_{n n'}^{\mathrm{G},- } = 0$. In the weak coupling or bad cavity limit, the electromagnetic modes can be traced out to obtain a reduced master equation for the QE only, as was done in Refs.~\cite{franke2019quantization, cirac1992interaction}, with the QE decay rate
\begin{align}\label{eqn:QE_decay_rate_1D}
    \Gamma_\mathrm{QE} = \frac{1}{\hbar^2} \Re\left( g_n P^{+\frac{1}{2}}_{n n'} \frac{1}{\mathrm{i}( \omega_\mathrm{QE} - \tilde{\omega}_{n'}^*)} P^{-\frac{1}{2}}_{n' n'' } g_{n''}^* \right) \ .
\end{align}
Normalizing this result with respect to the free-space decay rate in 1D, $\Gamma_0 = {\omega_\mathrm{QE}|d|^2/}{2\hbar n_\mathrm{b} \varepsilon_0 c A}$, yields the Purcell factor.
%This result can be used to calculate the Purcell factor with the 1D free space decay rate $\Gamma_0 = \frac{1}{2 n_\mathrm{b}} \frac{|d|^2}{\hbar \varepsilon_0 c A} \omega_\mathrm{QE}$.

For our calculations using Eq.~\eqref{eqn:QE_decay_rate_1D}, we retain the three lowest-order QNMs of the resonator which lie in spectral proximity to the QE transition $\omega_\mathrm{QE} = 1.95\,\mathrm{eV}$, with complex frequencies $\tilde{\omega}_1 = (0.65-0.18\mathrm{i})\,\mathrm{eV}$, $\tilde{\omega}_2 = (1.95-0.18\mathrm{i})\,\mathrm{eV}$ and $\tilde{\omega}_3 = (3.25-0.18\mathrm{i})\,\mathrm{eV}$, as shown in the upper panel of Fig.~\ref{fig:1D_Cavity}(b). These modes have very low quality factors of $Q_1 = 1.76$, $Q_2 = 5.28$ and $Q_3 = 8.80$.
%\textcolor{red}{Furthermore, we set the dimensionless dipole moment to be $|{\bm d}|/\sqrt{\hbar \varepsilon_0 c A} = 0.3$}, adequate for weak coupling, and
%$\tilde{\omega}_1 = (0.6491-0.1844\mathrm{i})\,\mathrm{eV}$, $\tilde{\omega}_2 = (1.9474-0.1844\mathrm{i})\,\mathrm{eV}$, $\tilde{\omega}_3 = (3.2457-0.1844\mathrm{i})\,\mathrm{eV}$
Furthermore, we assume that the QE is located at $x_\mathrm{QE} = 66\, \mathrm{nm}$, where an electric-field antinode of the second-order QNM occurs [see Fig.~\ref{fig:1D_Cavity}(a)]. Under these conditions, whilst all three QNMs of the resonator offer polarization matching, only the second-order QNM allows optimal spectral and spatial matching with the QE transition.
Fig.~\ref{fig:1D_Cavity}(b) presents our calculated Purcell factor, which bears the expected structure comprising three Lorentzian-like peaks at the spectral locations of the QNMs. In the weak coupling regime considered here, the decay rate can be derived semi-classically using Fermi's golden rule and the Green's function of the electric field~\cite{novotnyhechtbook}, yielding (in 1D) $\Gamma^{\textrm{SC}}_\mathrm{QE} = \omega_\mathrm{QE}^2|d|^2\Im(G(x_\mathrm{QE},x_\mathrm{QE},\omega_\mathrm{QE}))/\hbar \varepsilon_0 c^2 A$.
%We considered the first three QNMs with complex frequencies $\tilde{\omega}_1 = (0.6491-0.1844\mathrm{i})\,\mathrm{eV}$, $\tilde{\omega}_2 = (1.9474-0.1844\mathrm{i})\,\mathrm{eV}$, $\tilde{\omega}_3 = (3.2457-0.1844\mathrm{i})\,\mathrm{eV}$ and a QE located at $x_\mathrm{QE} = 66\, \mathrm{nm}$, which is a local intensity maximum of the second QNM (see Fig.~\ref{fig:1D_Cavity_1}(a)).
The predictions obtained using the fully quantum and semi-classical models are compared in the lower panel of Fig.~\ref{fig:1D_Cavity}(b), showing excellent agreement. Note that we have considered only a finite frequency interval $[\omega_\mathrm{min},\omega_\mathrm{max}]$ for the numerical integration in Eq.~\eqref{eqn:P_integral_1D}, with $\omega_\mathrm{min}= 0.25\,\mathrm{eV}$ and $\omega_\mathrm{max} = 4.0\,\mathrm{eV}$.

In the next step, we compare the electric field and QE dynamics in the strong coupling regime. In Ref.~\cite{dung2000mqed,scheel2008macroscopic}, the spontaneous decay of QEs was studied within the framework of mQED. If the QE is prepared in the excited state $\ket{\mathrm{e}}$ at time $t=0$, and the electromagnetic field in the vacuum state $\ket{0}$, then the time-dependent state vector of the system can be written as $\ket{\psi(t)} = C_\mathrm{e}(t) \mathrm{e}^{-\mathrm{i}\omega_\mathrm{QE} t} \ket{\mathrm{e}}\ket{0} + \int\mathrm{d}x\int\mathrm{d}\omega\,C_{\mathrm{g},\lambda}(x,\omega,t) \mathrm{e}^{-\mathrm{i} \omega  t }\ket{\mathrm{g}}\hat{f}_\lambda^\dagger(x,\omega)\ket{0}$. From the Heisenberg equations of mQED, an integral equation for the time-dependent coefficient $C_\mathrm{e}(t)$ can be derived:
\begin{align}\label{eqn:mQED_Integral_Eq}
    C_\mathrm{e}(t) = 1 + \int_0^\mathrm{t}\mathrm{d}\tau\, K(t-\tau) C_\mathrm{e}(\tau) \ , 
\end{align}
with the kernel
%\textcolor{red}{Check arguments of the Green's function here. Also, should the notation for the imaginary part be kept consistent throughout [c.f. Eq. (22)]?}
\begin{align}
    K(t-\tau) = \frac{|d|^2}{ \hbar \varepsilon_0 \pi A} \int_{\omega_\mathrm{min}}^{\omega_\mathrm{max}}\mathrm{d}\omega\, \frac{\omega^2}{c^2}\frac{\Im(G( x_\mathrm{QE},x_\mathrm{QE},\omega ))}{\mathrm{i}(\omega-\omega_\mathrm{QE})} \notag \\
    \cdot\left( \mathrm{e}^{-\mathrm{i}(\omega - \omega_\mathrm{QE})(t-\tau)} - 1  \right) \ . 
\end{align}
This equation is derived from the exact mQED formulation and involves no approximations beyond the rotating-wave one for the interaction with the QE. We have solved the integral equation numerically, considering a QE
%frequency $\omega_\mathrm{QE} = 1.9464\,\mathrm{eV}$ ($637\,\mathrm{nm}$), corresponding to a nitrogen-vacancy center~\cite{aharonovich2014diamond} and
tuned to the second-order QNM (\textit{i.e.}, $\omega_\mathrm{QE} = 1.95\,\mathrm{eV}$) and setting the dimensionless dipole moment to be $|d|/\sqrt{\hbar \varepsilon_0 c A} = 0.3$, which although unrealistically high, serves to demonstrate the validity of our theory in the strong coupling case. The result for the occupation probability of the excited state, $|C_\mathrm{e}(t)|^2$, can then be compared to the expectation value $\langle \sigma^+(t) \sigma^-(t)\rangle$, obtained from the quantum QNM master equation that we solve using the QuTiP library~\cite{lambert2024qutip}. The numerical data are plotted in Fig.~\ref{fig:1D_cavity_dynamics}(a), evidencing excellent agreement between the exact theory and the quantum QNM approach in capturing the temporal Rabi oscillations characteristic of this coupling regime. We have further plotted the expectation values of the QNM number operators $\langle \hat{a}^\dagger_n(t) \hat{a}_n(t) \rangle$; as expected, the symmetrized, second-order mode is dominant and undergoes coherent energy exchange with the QE, as reflected in the Rabi oscillations. Moreover, the amplitudes decrease in an approximately linear fashion on the logarithmic plot, with a gradient corresponding to the (equal) imaginary parts of the QNM frequencies $\Im(\tilde{\omega}_n) = - 0.184\,\mathrm{eV}$. After the first Rabi oscillation, the other two symmetrized modes show qualitatively similar oscillations as the second-order one, which might be explained by the contribution of the second-order QNM in the other symmetrized modes. Prior to this (\textit{i.e.}, during the first $\sim 10\,\mathrm{fs}$), their dynamics is more complicated and shows additional structure caused by the excitation of the first- and third-order QNMs. 

From the exact mQED formulation, one can further derive an equation for the expectation value of the electric field intensity emitted via the spontaneous decay of the QE~\cite{dung2000mqed}:
\begin{align}
    \langle \hat{E}^{(-)}(x,t)\hat{E}^{(+)}(x,t) \rangle = \bigg|\frac{d}{\pi \varepsilon_0 c^2 \sqrt{A}} \int_0^t\mathrm{d}\tau\, C_\mathrm{e}(\tau) \notag \\ 
    \cdot\int_{\omega_\mathrm{min}}^{\omega_\mathrm{max}} \mathrm{d}\omega\, \omega^2 
    \Im(G(x,x_\mathrm{QE},\omega_\mathrm{QE}) ) \mathrm{e}^{-\mathrm{i} (\omega-\omega_\mathrm{QE})(t-\tau) } \bigg|^2 \ . \label{eqn:I_out_mQED}
\end{align}
In Fig.~\ref{fig:1D_cavity_dynamics}(b), we compare the exact result for the normalized output intensity at $x = L$ with that obtained from the QNM master equation; once again, very good agreement can be seen. We observe two peaks of the intensity attributable to the Rabi oscillations of the second-order mode around $\sim 5\,\mathrm{fs}$ and $\sim 17\, \mathrm{fs}$. However, the first peak has additional substructure. Its maximum is split into a pair of maxima and the second derivative is changing sign along the drop of the peak. For comparison, we plot the intensity calculated from the quantum QNM model upon neglecting the first- and third-order modes (\textit{i.e.}, $ | \tilde{E}^\mathrm{sym}_{2}(x) |^2 \langle \hat{a}^\dagger_2(t) \hat{a}_2(t) \rangle$). From this, we can appreciate that the additional substructure originates from a multimode effect, as the intensity in this case only displays the two peaks from the Rabi oscillations. 

Collectively, our results show that the derived master equation correctly reproduces the dynamics of the QE and electric field, seeding confidence in its application to more complex, quantum nanophotonic systems.  

%\textcolor{red}{Also, the figures should be in reasonable proportion -- can Fig. 3(b) be increased in width without distortion?}

It is noteworthy that in this 1D example, we can place the PMLs directly at the open end of the resonator. Positioning the PMLs away from the open end would degrade the accuracy of the results, and increasingly so the farther away they are shifted, given the
%if only the QNMs were considered because of their
divergence and incompleteness of the QNMs outside the resonator. In such circumstances, a correct expansion of the Green's function in this region would necessitate taking the PML modes into account. For 3D systems, where there is a transition between the near-field and the divergent far-field regions, optimizing the separation between the resonator and PMLs should aid in reducing the required number of PML modes, a matter to be investigated in greater detail in future studies.
%This matter will be investigated in more detail in further studies.  

% \begin{figure}
%     \centering
%     \includegraphics[width=1.0\linewidth]{Fig_1D_cavity_combined_2.png}
%     \caption{Numerical exploration of a 1D, half-open, diamond cavity in air, with an electric-dipole QE embedded therein. (a) Schematic of the cavity-QE system with characteristic parameters indicated. The dashed line shows the electric field intensity profile of the second-order QNM, to which the QE is resonantly coupled. (b) Complex frequencies $\tilde{\omega}_n$ of the three, lowest-order QNMs (upper panel) and the frequency-dependent Purcell factor of a QE with $\omega_\mathrm{QE} = 1.95\,\mathrm{eV}$ ($637\,\mathrm{nm}$), calculated in accordance with the quantum QNM and semi-classical theories (lower panel). (c) QE and symmetrized mode dynamics (left panel) and (normalized) output electric field intensity at $x=L$ (right panel) for the spontaneous decay of a QE with $\omega_\mathrm{QE} = 1.95\,\mathrm{eV}$ and  $|d|/\sqrt{\hbar \varepsilon_0 c A} = 0.3$, obtained by solving the quantum QNM master equation and the integral equation Eq.~\eqref{eqn:mQED_Integral_Eq} in mQED. In the right panel, we have further plotted the electric field intensity calculated from the quantum QNM model retaining only the second-order mode $\hat{a}_2$. }
%     \label{fig:1D_Cavity_1}
% \end{figure}

\begin{figure}
    \centering
    \includegraphics[width=1.0\linewidth]{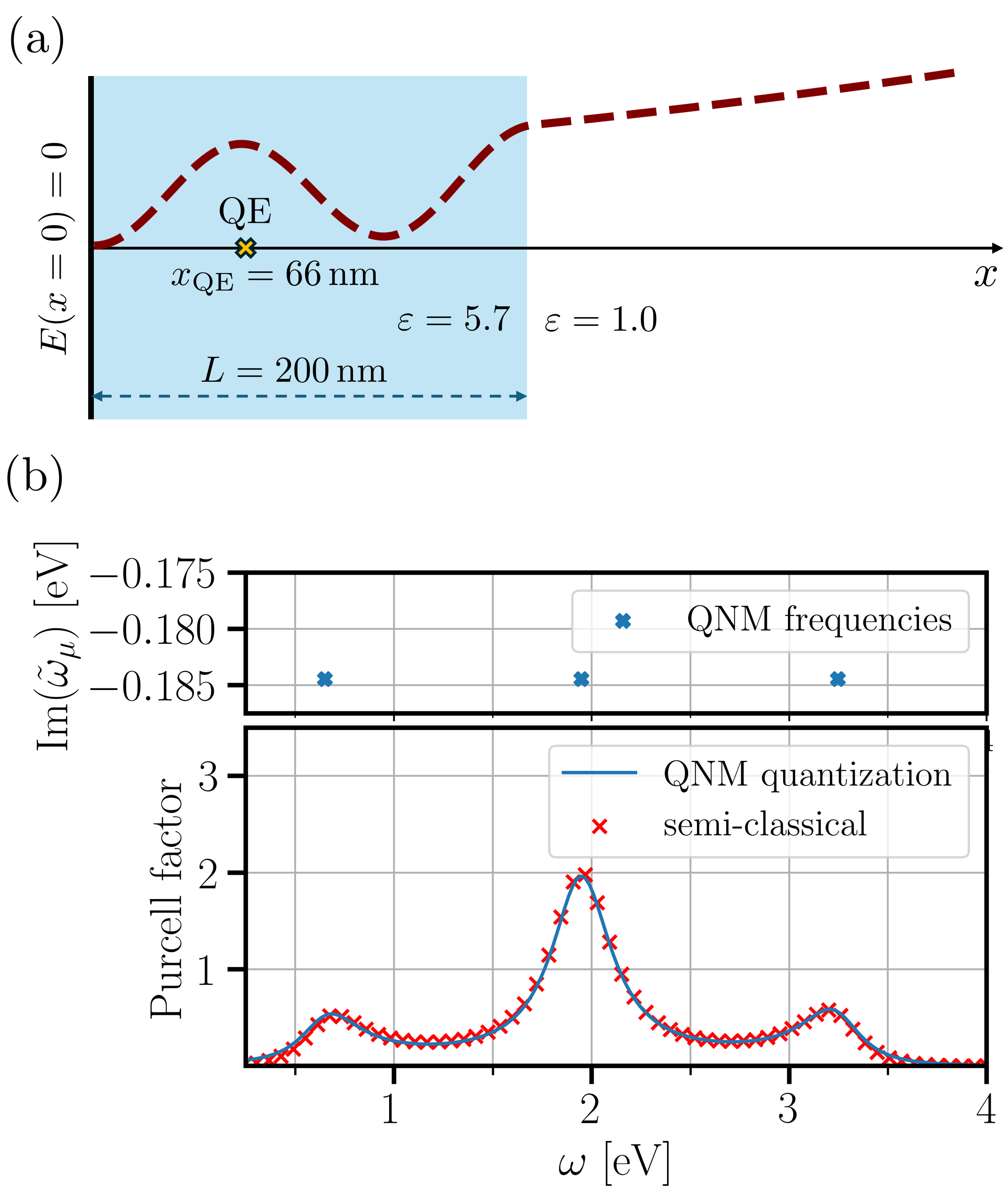}
    \caption{Numerical exploration of a 1D, half-open, diamond cavity in air, with an electric-dipole QE embedded therein. (a) Schematic of the cavity-QE system with characteristic parameters indicated. The dashed line shows the electric field intensity profile of the second-order QNM. (b) Complex frequencies $\tilde{\omega}_n$ of the three, lowest-order QNMs (upper panel) and the frequency-dependent Purcell factor of the QE, calculated in accordance with the quantum QNM and semi-classical theories (lower panel). }
    \label{fig:1D_Cavity}
\end{figure}

\begin{figure}
    \centering
    \includegraphics[width=1.0\linewidth]{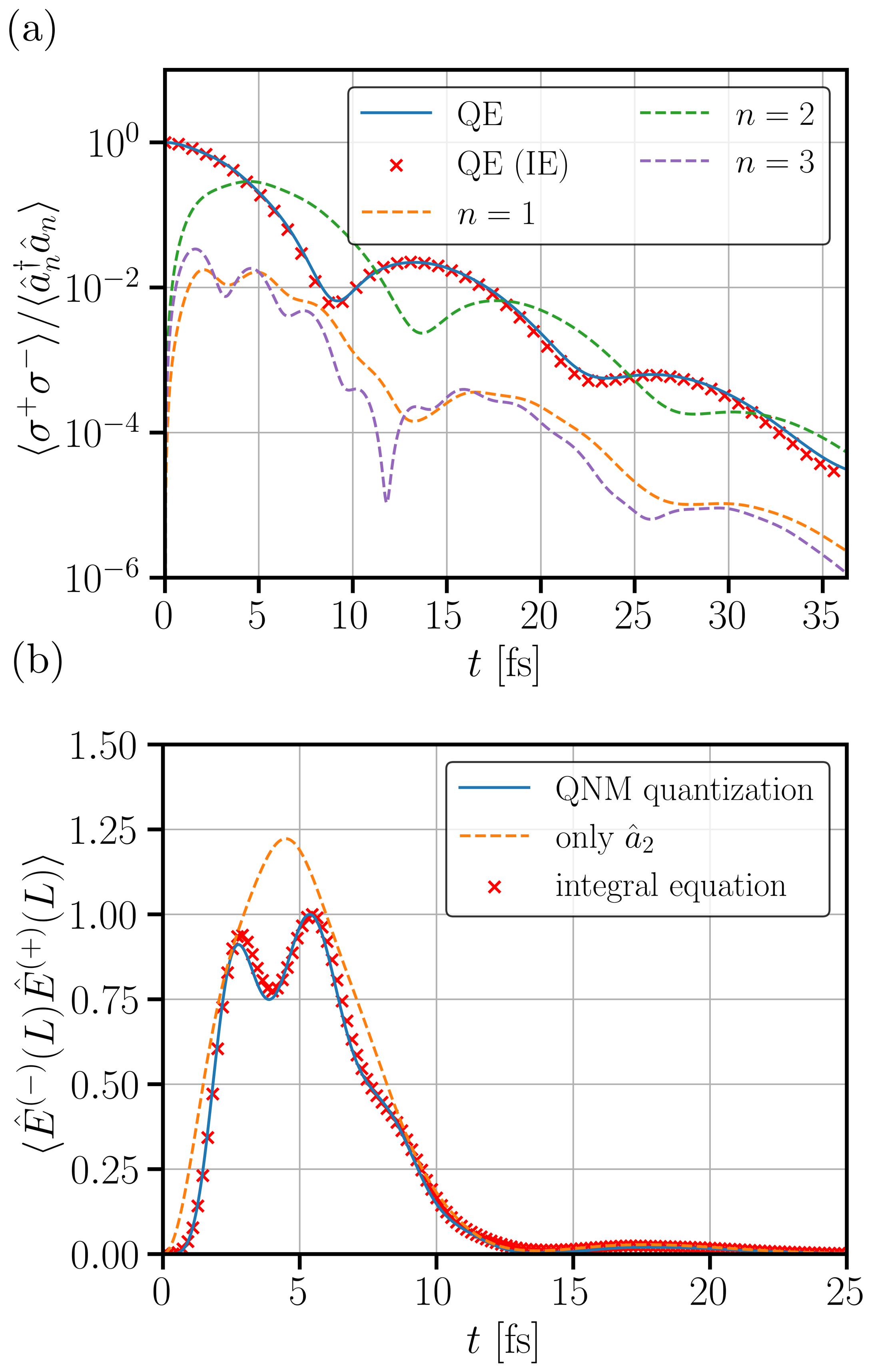}
    \caption{Spontaneous decay of a QE with $\omega_\mathrm{QE} = 1.95\,\mathrm{eV}$ and  $|d|/\sqrt{\hbar \varepsilon_0 c A} = 0.3$ in the 1D, half-open cavity of Fig.~\ref{fig:1D_Cavity}(a). (a) QE and symmetrized mode dynamics obtained by solving the QNM master equation, compared to the QE dynamics obtained from the integral equation Eq.~\eqref{eqn:mQED_Integral_Eq} in mQED. (b) Normalized output electric field intensity at $x = L$ obtained from the QNM master equation, compared to the exact mQED result from Eq.~\eqref{eqn:I_out_mQED}. Also shown is the electric field intensity calculated from the quantum QNM model retaining only the second-order mode $\hat{a}_2$. }
    \label{fig:1D_cavity_dynamics}
\end{figure}

\subsection{\label{Sec:3D_example} 3D Spherical Cavity}

To demonstrate the validity of our theory in the 3D case, we calculate the Purcell factor of a dipolar QE embedded in a spherical, dielectric resonator. Owing to the high degree of symmetry, both the QNMs and the $P$-matrix elements can be determined semi-analytically, just as in the 1D case, but with the crucial difference that in 3D, the PMLs include a gain contribution that must be correctly eliminated from the quantum dynamics, as discussed in \S~\ref{Sec:Quantum_QNMs}. Specifically, we consider a silicon (Si) sphere with $\varepsilon_\mathrm{Si} = 12$ and radius $R = 850\,\mathrm{nm}$ in vacuum, as shown in Fig.~\ref{fig:3D_Cavity}(a). In the presence of ideal spherical symmetry, the modes can be classified as transverse magnetic and transverse electric modes, $\mathrm{TM}_{l m n}$ and $\mathrm{TE}_{l m n}$ respectively, with the angular quantum numbers $(l, m)$ ($l \geq 1$, $|m| \leq l $) and the radial quantum number $n$ ($n = 1,2,\dots$)~\cite{strattonbook}. Furthermore, TE and TM modes with different $l$ or $m$ do not pair in the $P$-matrix due to the orthogonality of the corresponding vector spherical harmonics, allowing for a separate treatment of each set of modes. As an illustration and for the sake of simplicity, we therefore consider three $\mathrm{TM}_{l=1}$ modes with radial quantum numbers $n = 4,5,6$ (each with a three-fold degeneracy) and complex frequencies $\tilde{\omega}_{1 m 4} = ( 0.94 - 0.02\mathrm{i} )\,\mathrm{eV}$, $\tilde{\omega}_{1 m 5} = ( 1.15 - 0.02\mathrm{i} )\,\mathrm{eV}$ and $\tilde{\omega}_{1 m 6} = ( 1.37 - 0.02\mathrm{i} )\,\mathrm{eV}$, which are plotted in the upper panel of Fig.~\ref{fig:3D_Cavity}(b). The contributions of these modes to the Purcell factor can be calculated as in Eq.~\eqref{eqn:QE_decay_rate_1D}, and as the analytic form of the Green's function for such a system is decomposed into $\mathrm{TM}_{l m}$ and $\mathrm{TE}_{l m}$ parts~\cite{li1994electromagnetic}, we can compare the contribution of these modes directly with the exact $\mathrm{TM}_{l = 1}$ contribution. The results for two different QE positions and orthogonal dipole orientations are shown in the lower panel of Fig.~\ref{fig:3D_Cavity}(b). We observe that the same resonances can have very different shapes for the two different dipoles, ranging from a Lorentzian-like peak to a strong Fano-like suppression of the decay rate, which the quantum QNM theory correctly reproduces to a high degree. The small deviation for the lowest-energy resonance likely stems from the contribution of other $\mathrm{TM}_{l=1}$ modes with lower radial quantum numbers $n$.

\begin{figure}
    \centering
    \includegraphics[width=1.0\linewidth]{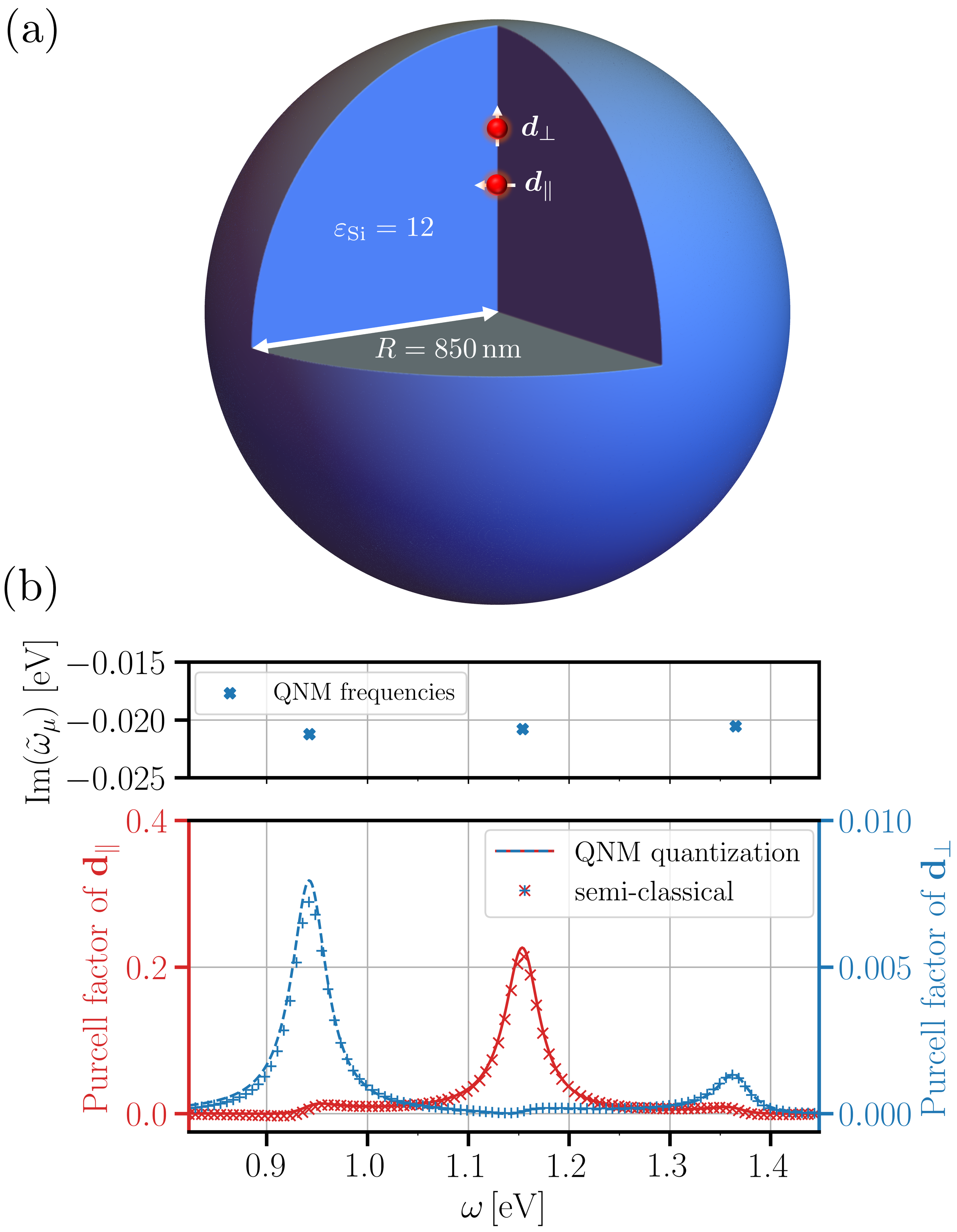}
    \caption{Numerical exploration of a 3D, spherical, Si cavity in air, with an electric-dipole QE embedded therein. (a) Schematic of the cavity-QE system with characteristic parameters indicated and two distinct coupling configurations, featuring a QE located at a radial distance of $r_{\bm{d}_\parallel} = 380\,\mathrm{nm}$ or $r_{\bm{d}_\perp} = 545\,\mathrm{nm}$ and with orthogonal dipole orientations. (b) Complex frequencies $\tilde{\omega}_{l m n}$ of the three considered $\mathrm{TM}_{l = 1}$ QNMs (upper panel) and the frequency-dependent contributions of these modes to the Purcell factor of the QE in each configuration, calculated in accordance with the quantum QNM and semi-classical theories (lower panel).}
    \label{fig:3D_Cavity}
\end{figure}

\section{\label{Sec:Conclusion} Conclusion}

In conclusion, we have introduced a formalism for the quantization of QNMs
%\textcolor{red}{(and additional PML modes)}
in the presence of 3D, spatially inhomogeneous, dissipative (with possible gain), linear magnetodielectric media, enabling a unified and rigorous approach for simulating photonic, plasmonic and magnonic cQED phenomena. Similar to a number of recent works~\cite{franke2019quantization,medina2021few,feist2020macroscopic}, our approach is rooted in the phenomenological Green's function quantization scheme of mQED,
%but offers the crucial advantage of a rigorously defined mode regularization via PMLs.
but relies on the use of PMLs to achieve a set of modes that are rigorously regularized, orthogonalized and complete, rendering them suitable for the representation of the dyadic Green's function and electromagnetic fields both inside and outside the resonator volume. 
The introduction of PMLs transforms the radiative losses into non-radiative material dissipation, and via a suitable transformation that reflects all the losses of the resonator, we are able to define creation and annihilation operators that allow the construction of modal Fock states for the joint excitations of field-dressed matter. Our quantized description of the QNMs thereby provides the foundations for a quantum dynamic treatment of magnetodielectric cQED systems, utilizing discrete modes together with rigorously derived quantum Langevin and master equations.
%Notably, the availability of an efficient implementation of the auxiliary-field eigenvalue approach~\cite{sauvan2013theory}, together with the widespread use of PMLs, render our approach naturally compatible with computational practice in nanophotonics. Additionally, it bears the advantage of avoiding more elaborate techniques for obtaining regularized QNMs and the dyadic Green's function outside the resonator, such as those based on a Dyson equation formalism~\cite{ge2014quasinormal,franke2019quantization}.
By directly addressing the intricacies of modal loss in a fully quantum theory of magnetodielectric cQED, our approach enables a proper assessment of the performance metrics for nanophotonic device protocols in the Quantum 2.0 era, extending from on-demand sources of single photons, plasmons and magnons, dynamic entanglement generation and quantum state transduction, to light-matter strong coupling in hybrid configurations. Moreover, our quantum QNM theory could prove especially valuable in revealing the opportunities that dissipation may offer in manipulating and tailoring the dynamics of quantum systems (contrary to its traditionally perceived adversarial role), such as in the development of quantum technologies based on
%near-field-driven single-photon sources~\cite{hoang2016ultrafast,hughes2019spe,bello2020controlled,dowran2023plasmon} and dynamic entanglement generation~\cite{yang2013quantum,xiong2020ultra,bello2020controlled,bello2022near}, to the real-time control of strong coupling~\cite{crai2020electron,yang2024switch,abad2024electron} and quantum biosensing~\cite{kongsuwan2019sensing}, as well as schemes that harness dissipation in a fundamentally instrumental way, such as the recently proposed
quantum nanoplasmonic coherent perfect absorption~\cite{lai2024room}, non-hermitian topological magnonics~\cite{yu2024mag} and dissipative photonic time crystals~\cite{lyubarov2022ptc,liu2023ctc,osuna2025pair}. Note that although the modification of mQED in the case of time-dependent material properties is not trivial, important first steps have very recently been made with the proposition of an mQED Lagrangian for a time-dependent Drude model~\cite{horsley2025macroscopic}, following the quantization of the electromagnetic field in lossless systems~\cite{pendry2024qed}.

\begin{acknowledgments}
The authors gratefully acknowledge funding from Research Ireland (formerly Science Foundation Ireland) via Grants No. 18/RP/6236 and 22/QERA/3821.
\end{acknowledgments}

\section*{Conflict of Interest}
The authors have no conflicts to disclose.

\section*{Data Availability}
The data that support the findings of this study are available from the corresponding authors upon reasonable request.

\nocite{*}
\bibliography{aipsamp}% Produces the bibliography via BibTeX.

\end{document}